%% file: main.tex
\documentclass[12pt]{iopart}
\usepackage{booktabs}
\usepackage{comment,mfirstuc}
\usepackage{graphicx}% Include figure files
\usepackage{grffile}
\usepackage{algorithm}% http://ctan.org/pkg/algorithm
\usepackage{algpseudocode}% http://ctan.org/pkg/algorithmicx
\usepackage{hyperref}% add hypertext capabilities
\usepackage{soul}
\usepackage{listings}
\usepackage{multirow}
\usepackage{dcolumn}% Align table columns on decimal point
\usepackage{bm}% bold math
\usepackage{placeins}
\expandafter\let\csname equation*\endcsname\relax
\expandafter\let\csname endequation*\endcsname\relax
\usepackage{amsmath}
\usepackage{amssymb}
\usepackage{mathtools}
\usepackage[bottom=1.0in]{geometry}
\usepackage{cite}
\usepackage{lineno}
\RequirePackage{xcolor}
\usepackage[misc]{ifsym}
\usepackage{lipsum} 
\usepackage{subcaption}
\usepackage{tcolorbox} % Create coloured boxes (e.g. the one for the key-words)
\usepackage{stfloats} % Correct position of the tables
\usepackage{siunitx}
\DeclareSIUnit\eVperc{\eV\per\clight}
\DeclareSIUnit\clight{\text{\ensuremath{c}}}

\usepackage{ulem}

\usepackage{xspace}
\usepackage{colortbl}
\usepackage{array}
\newcolumntype{P}[1]{>{\centering\arraybackslash}p{#1}}
\newcolumntype{M}[1]{>{\centering\arraybackslash}m{#1}}
\hypersetup{ 
    pdfnewwindow=true,      % links in new window
    colorlinks=true,       % false: boxed links; true: colored links
    linkcolor=blue,         % color of internal links
    citecolor=blue,        % color of links to bibliography
    filecolor=blue,      % color of file links
    urlcolor=blue        % color of external links
}  

\algblock{Input}{EndInput}
\algnotext{EndInput}
\algblock{Output}{EndOutput}
\algnotext{EndOutput}

\def\be{\begin{eqnarray} &&} 
 
\def\ee{\end{eqnarray}}

\makeatletter
\newcommand{\mainmatter}{%
  \setcounter{footnote}{0}%
  \patchcmd{\@makefntext}{\fnsymbol}{\arabic}{}{}%
  \patchcmd{\@thefnmark}{\fnsymbol}{\arabic}{}{}%
  \def\@makefnmark{\textsuperscript{\arabic{footnote}}}
  \long\def\@makefntext##1{\parindent 1em\noindent
        \hb@xt@1.8em{%
            \hss\@textsuperscript{\normalfont\@thefnmark}}##1}%
}
\makeatother

\newcommand{\addComment}[2]{
  \expandafter\newcommand\csname #1\endcsname[1]{{\bf \color{#2} \capitalisewords{#1}:\,##1}}
  \expandafter\newcommand\csname #1cor\endcsname[2]{{\color{#2} \capitalisewords{#1}:\,\st{##1}{\bf ##2}}}
  \expandafter\newcommand\csname #1color\endcsname{#2}
}
\addComment{cris}{blue} 
\addComment{james}{red}
\addComment{cole}{purple}

\newcommand{\gluex}{\textsc{GlueX}\xspace} 
\newcommand{\geant}{\textsc{Geant4}\xspace}

\begin{document}

% Keywords command
\providecommand{\keywords}[1]
{
  \small
  \textbf{Keywords:}  {\color{blue}#1
  }
}

\title[\scriptsize{Application of a Mixture of Experts-based Foundation Model to the GlueX DIRC Detector}]{Application of a Mixture of Experts-based Foundation Model to the GlueX DIRC Detector} 

\author{C. Fanelli$^{1,2,\star}$, J. Giroux$^{1,\star}$, C. Granger$^{1,\star}$, J. Stevens$^{2,\star}$} 

\address{
$^{1}$ William \& Mary, Department of Data Science, Williamsburg, VA 23185, USA\\
$^{2}$ William \& Mary, Department of Physics, Williamsburg, VA 23185, USA\\
$^{\star}$ Author to whom any correspondence should be addressed.
}

\ead{{\color{blue}
\{cfanelli,
jgiroux,
cjgranger,
jrstevens01\}@wm.edu
}}

\vspace{10pt}
\begin{indented}
\item[]\today
\end{indented}

%\linenumbers
\begin{abstract}

We present a Mixture-of-Experts-based foundation model applied to the \gluex DIRC detector at Jefferson Lab, demonstrating its utility as a unified framework for fast simulation, particle identification, and hit-level noise filtering of Cherenkov photons. By leveraging a single shared transformer backbone across all tasks, the approach eliminates the fragmentation of task-specific pipelines while maintaining competitive—and in several cases superior—performance relative to established methods. The model operates directly on low-level detector inputs, performing hit-by-hit autoregressive generation over split spatial and temporal vocabularies with continuous kinematic conditioning, and supports class-conditional generation of pions and kaons through its Mixture-of-Experts architecture. We benchmark against the standard geometrical reconstruction and prior deep learning methods across the full kinematic phase space of the \gluex DIRC, demonstrating that the foundation model framework transfers effectively to this detector without architectural modification. This work positions the foundation model as a practical and scalable alternative to the suite of task-specific models currently proposed for \gluex DIRC analysis.

\end{abstract}


%\maketitle

\mainmatter

\input{1_introduction}

\input{2_data}

\input{3_architecture}

\input{4_analysis}

%\input{5_impacts}

%\clearpage

\input{6_summary}

\section*{Data availability statement}
The data that support the findings of this study are available upon reasonable request from the authors.

\section*{Code Availability}
The code is publicly available at \href{https://github.com/wmdataphys/GlueX_DIRC_FM}{https://github.com/wmdataphys/GlueX\_DIRC\_FM}.

%\clearpage

\section*{Acknowledgments}
%\james{Cris check these.}
%
%We thank William \& Mary for supporting the work of JG and CF through CF's start-up funding. 
%
This material is based upon work supported by the National Science Foundation under Grant No. 2443510.
The authors acknowledge William \& Mary  Research Computing for providing computational resources and technical support that have contributed to the results reported within this article.
%

%\clearpage

\section*{References}
\bibliographystyle{iopart-num}
\bibliography{biblio}

\clearpage
\input{Appendix}

\end{document}

%% file: 1_introduction.tex
\section{Introduction}\label{sec:intro}

The identification of charged hadrons is a cornerstone of experimental nuclear and particle physics, underpinning measurements that probe the fundamental structure of matter. In experiments such as \gluex at Jefferson Lab~\cite{adhikari2021gluex}, this capability relies critically on Detection of Internally Reflected Cherenkov (DIRC) light detectors, which exploit total internal reflection within fused silica radiators to preserve and image the angular information of Cherenkov radiation at the photon level. The resulting hit patterns encode rich information about the particle species, but their complexity poses significant challenges for both real-time identification and high-fidelity simulation at scale.

Traditionally, particle identification (PID) in the \gluex DIRC has relied on geometrical reconstruction via a look-up-table (LUT), which computes likelihood differences between mass hypotheses by backpropagating observed hits through the known detector optics~\cite{stevens2016gluex,patsyuk2018status}. While effective at low momenta, this approach degrades as the Cherenkov angles of pions and kaons converge above $\sim 3\,\text{GeV}/c$, and its dependence on idealized geometric assumptions limits its fidelity in complex kinematic regions. On the simulation side, full \geant \cite{GEANT4:2002} tracking of Cherenkov photons through the optical system is computationally expensive, motivating the development of fast simulation surrogates that can reproduce detector response at a fraction of the cost.

Deep learning has emerged as a natural tool for both problems. Methods ranging from convolutional architectures~\cite{fanelli2020deeprich} to hierarchical vision transformers~\cite{fanelli2024deep,Liu_2021} and normalizing flows (NF)~\cite{fanelli2024deep,giroux2025generativemodelsfastsimulation,papamakarios2021normalizingflowsprobabilisticmodeling} have demonstrated strong performance on DIRC data, each advancing the state of the art in their respective tasks. Yet these approaches share a common limitation: each downstream task — classification, simulation, noise filtering — often demands a specialized model with no shared architectural structure. This fragmentation carries real costs in terms of training overhead, deployment complexity, and the inability to leverage complementary information across tasks.

The emergence of Foundation Models (FM) in physics~\cite{giroux2025towards,birk2024omnijet,Mikuni_2025,birk2025omnijet,hsu2026evenetfoundationmodelparticle,elsharkawy2026omnimoltransferringparticlephysics,mikuni2025omnicosmostransferringparticlephysics,park2025fm4nppscalingfoundationmodel,finke2023learning,bardhan2025hepjepa,leigh2025tokenization,vigl2024finetuning,Golling_2024,wildridge2024bumblebee,harris2024re,Butter2025} offers a path toward unification, with recent work demonstrating that a single transformer backbone can support diverse downstream tasks through lightweight task-specific heads. In~\cite{giroux2025towards}, the authors propose a FM tailored to the specific demands of low-level imaging Cherenkov data: hit-by-hit autoregressive generation over split spatial and temporal vocabularies, continuous kinematic conditioning, class-conditional generation via a Mixture-of-Experts (MoE) \cite{shazeer2017outrageously} architecture, and unified multi-task operation across generation, PID, and noise filtering. This was demonstrated on the hpDIRC at the future EIC~\cite{kalicy2024high}, where the model achieved strong performance across all tasks within a single unified framework.

In what follows, we apply the model of~\cite{giroux2025towards} directly to the \gluex DIRC, evaluating its performance across fast simulation, particle identification, and hit-level noise filtering, benchmarking against both geometrical reconstruction and prior deep learning methods. Beyond serving as a transferability study, this work positions the foundation model as a practical, unified alternative to the collection of task-specific models currently proposed for \gluex DIRC analysis.

%% file: 2_data.tex
\section{Dataset}\label{sec:data}

\begin{figure}[H]
    \centering 
    \includegraphics[width=\textwidth]{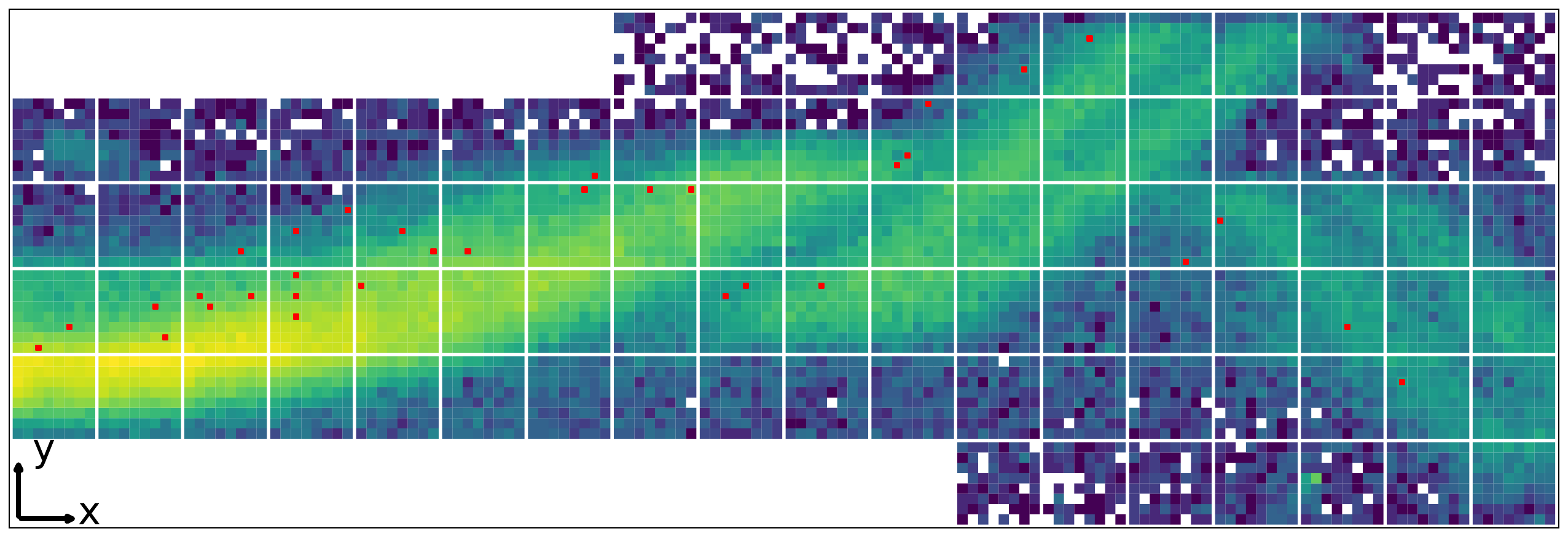} \\
    \caption{\textbf{\gluex DIRC Readout:} Individual charged particle tracks produce sparse hit patterns on the \gluex DIRC readout plane (red points), while the underlying probability density function (PDF) is recovered by accumulating many tracks with identical kinematics. White regions correspond to PMT locations absent from the detector due to low photon accumulation. 
    }
    \label{fig:sparse_hits}
\end{figure}

We utilize the dataset developed in~\cite{fanelli2024deep}, which records Cherenkov photon hit patterns captured by the \gluex DIRC, which operates in the forward region of the \gluex experiment at Jefferson Lab. The \gluex DIRC consists of 48 fused silica bars segmented into four bar boxes, and two readout zones (optical boxes) filled with distilled water and equipped with highly reflective mirrors that direct Cherenkov photons to arrays of Photomultiplier Tubes (PMTs). Each PMT contains 64 sensors of $6\times \SI{6}{\milli\meter}^2$ arranged in an $8 \times 8$ grid, with the PMTs themselves organized into a $6 \times 18$ array as depicted in Figure~\ref{fig:sparse_hits}. White spaces represent regions where PMTs have not been installed due to low accumulation of hits. For each detected photon, the readout provides the spatial location of the hit on the PMT plane together with a time-of-arrival measurement, providing a three-dimensional $(x, y, t)$ representation.

Each hit pattern is associated with a single charged particle track, characterized by its kinematic parameters. As depicted in Figure~\ref{fig:sparse_hits}, an individual track produces a sparse and inherently stochastic subset of hits, whereas accumulating hits from multiple tracks with identical kinematics reveals a well-defined underlying probability density function (PDF). The number of hits per track is not fixed and varies with kinematics, giving rise to a combinatorially large space of possible hit configurations---a significant challenge for generative and autoregressive architectures alike. The data, produced using full \textsc{Geant4} simulations via the \gluex software framework, correspond to charged pions ($\pi$) and kaons ($K$) distributed approximately uniformly over the detector acceptance, spanning $0.5 < |\vec{p}| < 6.5~\text{GeV}\,c^{-1}$, $0 < \theta < 11^{\circ}$, and $0 < \phi < 360^{\circ}$. We limit ourselves to a maximum sequence length of $180$ hits per track, accommodating the wide variation in photon yields across the phase space. This is a generous upper bound on the photon yield given that most tracks correspond to $\sim 50$ photons on average.

%% file: 3_architecture.tex
\section{Methods}\label{sec:methods} 

We deploy the foundation model architecture introduced in~\cite{giroux2025towards} directly to the \gluex DIRC, requiring no significant architectural modifications. We briefly summarize the model here for completeness; the reader is referred to~\cite{giroux2025towards} for full details.

\subsection*{Tokenization}

The pixelated nature of the \gluex DIRC readout naturally lends itself to tokenization: each discrete pixel index maps uniquely to a fixed $(x, y)$ coordinate in the PMT array, yielding a spatial vocabulary of size 5670 when accounting for removed PMTs. Time, a continuous variable, is discretized such that the vocabulary sizes are approximatley equal. Specifically, we span a range of $20-\SI{350}{\nano\second}$ in bin widths of $\SI{0.06}{\nano\second}$.

\subsection*{Architecture}

The model operates on two parallel sequences---spatial and temporal---each independently embedded via learnable projections from their respective vocabularies, with positional encodings added along each dimension. Kinematic conditioning is achieved by projecting the momentum magnitude $|\vec{p}|$, polar angle $\theta$ and azimuthal angle $\phi$ from continuous space and prepending them as contextual tokens to both sequences:

\begin{align}\label{eq:tokens}
    \text{spatial} &\rightarrow \{|\vec{p}|,\,\theta, \, \phi, \,\texttt{SOS}_p,\,p_1, \ldots, p_{n},\,
    \texttt{EOS}_p\} \notag \\
    \text{time}    &\rightarrow \{|\vec{p}|,\,\theta,\, \phi, \,\texttt{SOS}_t,\,t_1, \ldots, t_{n},\,
    \texttt{EOS}_t\}
\end{align}

Spatio-temporal information is fused through a Causal Multi-Head Cross-Attention (CMHCA) block, in which the Query (Q) projection is derived from the time embeddings and the Key (K) and Value (V) projections from the spatial embeddings---encoding the physical intuition that, for a given photon arrival time, one queries the set of geometrically valid pixel locations. The resulting spatio-temporally aware representations are passed through standard transformer blocks employing Multi-Head Self-Attention (MHSA)~\cite{vaswani2017attention}. Each block uses 8 attention heads, a linear projection layer, and a feed-forward neural network (FFNN) with GeLU activation~\cite{hendrycks2016gaussian}. A pre-normalization scheme~\cite{ba2016layer,xiong2020layer} is applied throughout, and $\ell_2$ normalization with a learnable scale factor is applied to the Q and K matrices~\cite{henry2020query}, promoting stable and direction-focused attention maps over the large combinatorial space of plausible track configurations. All embedding dimensions are set to 256.

To handle conditional generation of individual particle types — pions and kaons — we employ a MoE~\cite{shazeer2017outrageously} architecture with fixed routing. We allocate 2 experts per particle type (4 total), routing inputs on a per-class basis. To encourage balanced utilization within each class's expert pair, we apply an auxiliary load-balancing loss enforcing uniform per-token expert usage. As in \cite{giroux2025towards}, when transitioning to tasks such as particle identification where only a single FFNN is used within the transformer blocks, we initialize the weights of this network as the average of all experts.

The foundation model supports multiple downstream tasks through lightweight task-specific heads appended to the shared transformer backbone. Namely, (\textit{i}) \textit{Generation:} two independent linear heads operate over the spatial and temporal vocabularies, producing next-token predictions autoregressively. The training loss is the sum of the cross-entropy (CE) losses over both vocabularies. (\textit{ii}) \textit{Particle Identification:} a \textit{CLS} token is prepended to each input sequence and its final latent representation is passed through a linear classification head, trained with binary cross-entropy (BCE) loss. (\textit{iii}) \textit{Hit Filtering:} per-token latent representations are passed through a linear layer operating element-wise to produce per-hit class scores. Given the high signal-to-noise ratio and therefore, class imbalance, training uses focal loss~\cite{lin2017focal} with $\alpha = 0.75$ and $\gamma = 2$.

%% file: 4_analysis.tex
\section{Analysis and Results}\label{sec:results}

\subsection*{Particle Identification}

Following the procedure outlined in~\cite{fanelli2024deep}, we utilize the same standard metrics, namely pion rejection and kaon efficiency, with the associated Area Under the Curve (AUC) as the primary figure of merit, where a perfect PID method yields an AUC of 1.0. We also directly compare performance of the FM to the methods developed in \cite{fanelli2024deep}, namely, the Swin Transformer and Delta-log-likelihood (DLL) method using NF (indicated NF-DLL in plots that follow).

Given the relatively low number of training samples ($\sim 700\text{k}$ of each class), we found that the FM was prone to overfitting. As a result, we implement a simple data augmentation strategy in which we probabilistically move pixels into adjacent locations.
{This perturbation is confined to the originating PMT, mitigating the introduction of large systematic biases in relation to the underlying PDF.
In addition, we also smear the timing of individual hits with a uniform distribution of $\pm \SI{1}{\nano\second}$. While simple, these augmentation strategies prove to be highly effective.

Figure~\ref{fig:pgun_results} shows the performance of each method integrated over the entire phase space (left), with the AUC indicated in the legend. The foundation model (FM), Swin Transformer, and NF-DLL method are each compared against the established geometrical reconstruction baseline. Note that for the FM, we have performed a full fine tuning from the next-token model's base weights. As in~\cite{fanelli2024deep}, the AUC as a function of incoming track momentum (in $500\,\text{MeV}$ bins) is shown in the right panel of Figure~\ref{fig:pgun_results}, with uncertainties estimated via bootstrapping. We estimate the uncertainties on both rejection and efficiency using Eq.~\ref{eq:uncertainty}, where $f$ denotes the desired metric and $N$ the corresponding sample size:

\begin{equation}\label{eq:uncertainty}
    \sigma_f = \sqrt{\frac{f(1-f)}{N}}
\end{equation}

This enables sampling of different efficiency-vs-rejection curves, from which we report the mean AUC and associated error as the 95\% quantiles over each momentum bin.

\begin{figure}[!]
    \centering
    \includegraphics[width=0.49\textwidth]{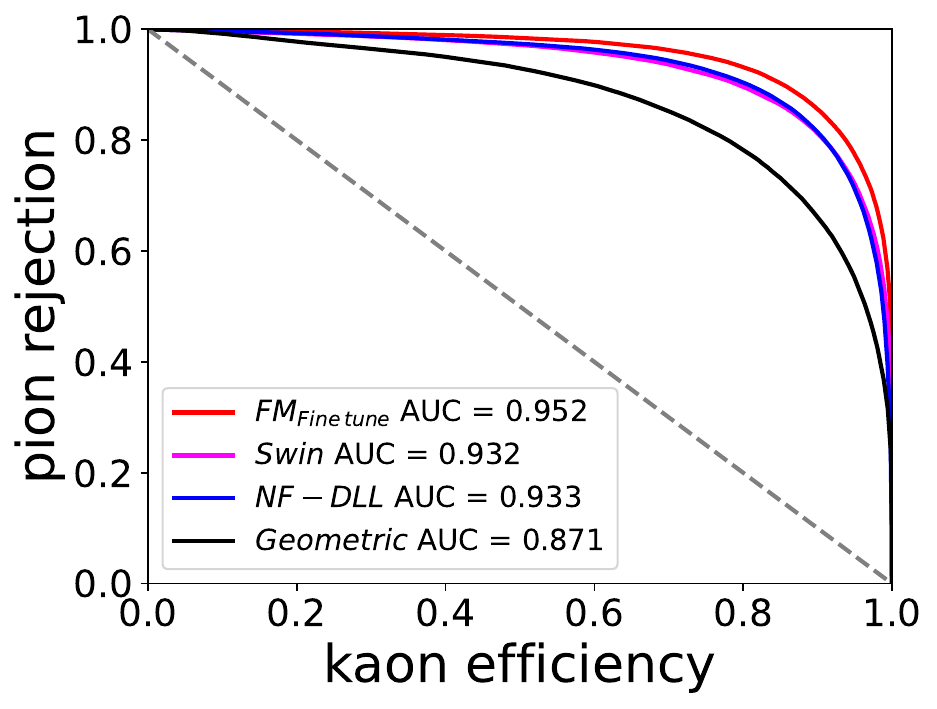}
    \includegraphics[width=0.48\textwidth]{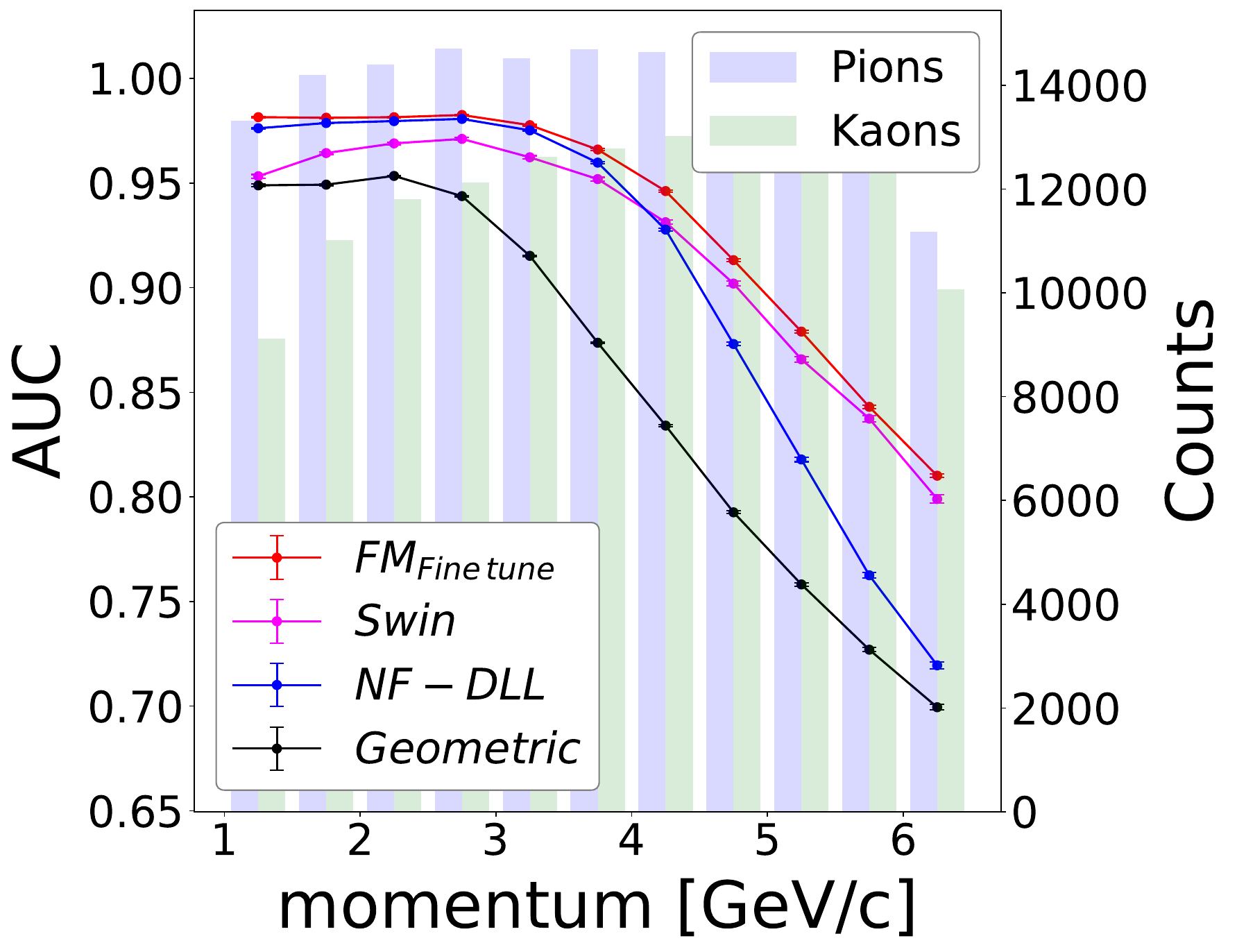}
    \caption{\textbf{Particle gun performance:} Pion rejection as a function of kaon efficiency 
    for the foundation model (FM), Swin Transformer, NF-DLL, and geometrical reconstruction, 
    integrated over the entire phase space (left). The AUC is indicated in the legend. AUC as a 
    function of track momentum (right), with uncertainty represented as 95\% quantiles obtained 
    via bootstrapping. Pion and kaon test counts are shown on the secondary axis (right).}
    \label{fig:pgun_results}
\end{figure}

From inspection of Figure~\ref{fig:pgun_results}, the foundation model achieves the highest AUC of $0.952$ integrated over the phase space, outperforming the Swin Transformer ($0.932$), NF-DLL ($0.933$), and the geometrical baseline ($0.871$). This performance advantage is consistent across all momentum bins, with all learned methods converging to similar performance at low momenta (below $\sim 3\,\text{GeV}/c$), after which the foundation model maintains the strongest discrimination as classification becomes increasingly difficult due to the Cherenkov angle of pions and kaons converging.
All learned methods significantly outperform geometrical reconstruction across the full momentum range, with the performance gap widening at higher momenta where the geometric approach degrades most rapidly.
%The NF-DLL method is known to be sensitive to noise in the data; with real experimental data and larger sample volumes,
%
We anticipate the foundation model to be the superior approach for pion-kaon classification in the \gluex DIRC.

%%%%%%%%%%%%%%%%%%%%%%%%%%% Generative Modeling %%%%%%%%%%%%%%%%%%%%%%%
\subsection*{Generative Model Evaluation}

We follow a similar procedure to that of~\cite{fanelli2024deep} to evaluate the generative quality of our foundation model across phase space. Specifically, we isolate specific bars in the DIRC alongside regions in $X$ along the face of the bar, providing kinematic constraints on $\theta$ and $\phi$ within a region and allowing integration over momentum space to visualize the Cherenkov rings from contributing angles. We have produced fast simulations across the entire phase space and select regions exhibiting the most complex ring structures for visualization.

Figure~\ref{fig:generations} shows pion (left column) and kaon (right column) generations for $X \in (0\,\text{cm}, 10\,\text{cm})$ at bar 10 (top row) and bar 31 (bottom row). These regions are centrally located within the DIRC and provide coverage of both optical boxes. Fast-simulated hit patterns are shown alongside their Geant4 ground truth counterparts.

\begin{figure}[!]
    \centering
    \includegraphics[width=0.49\textwidth]{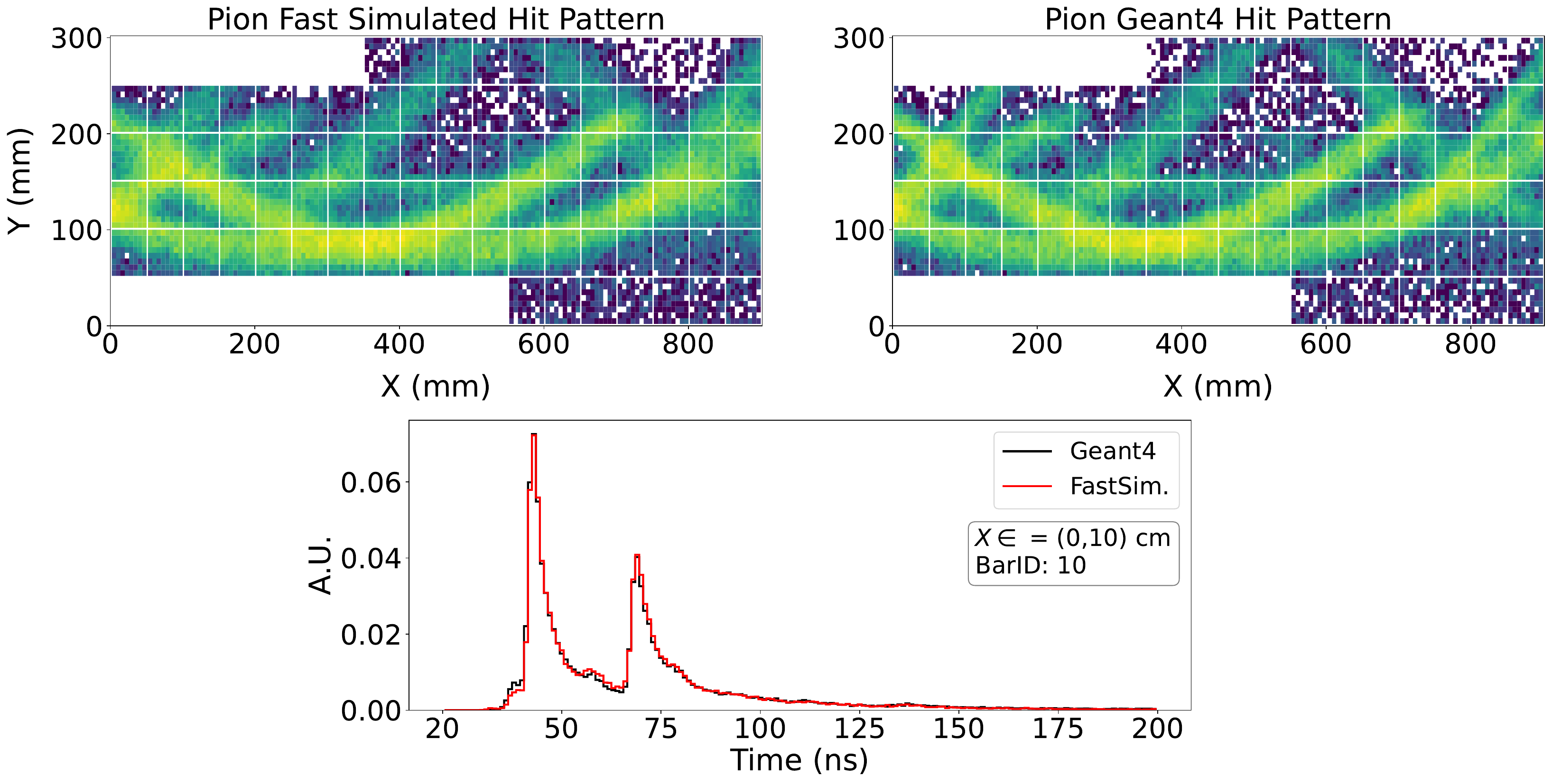}
    \includegraphics[width=0.49\textwidth]{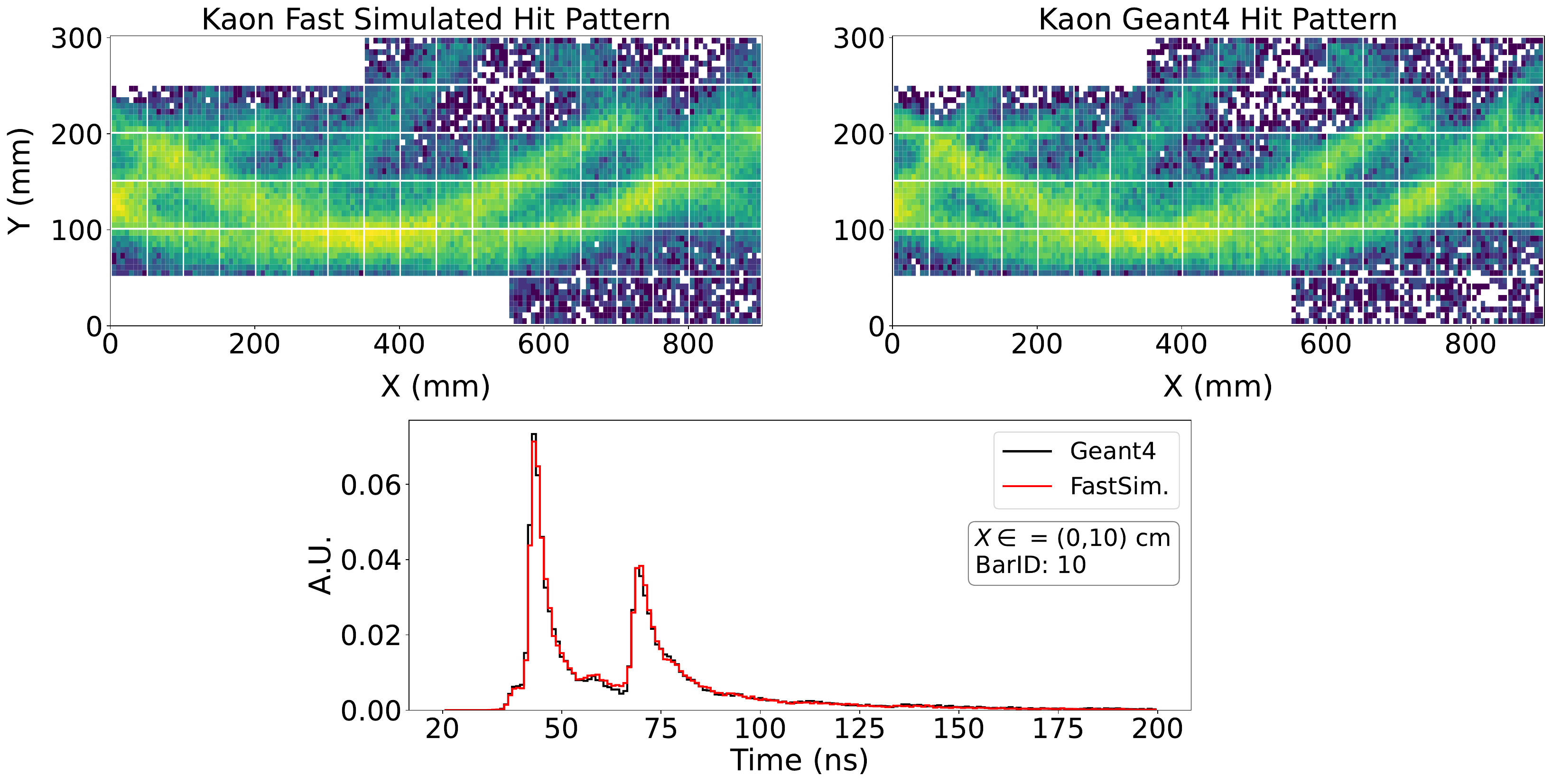}
    \includegraphics[width=0.49\textwidth]{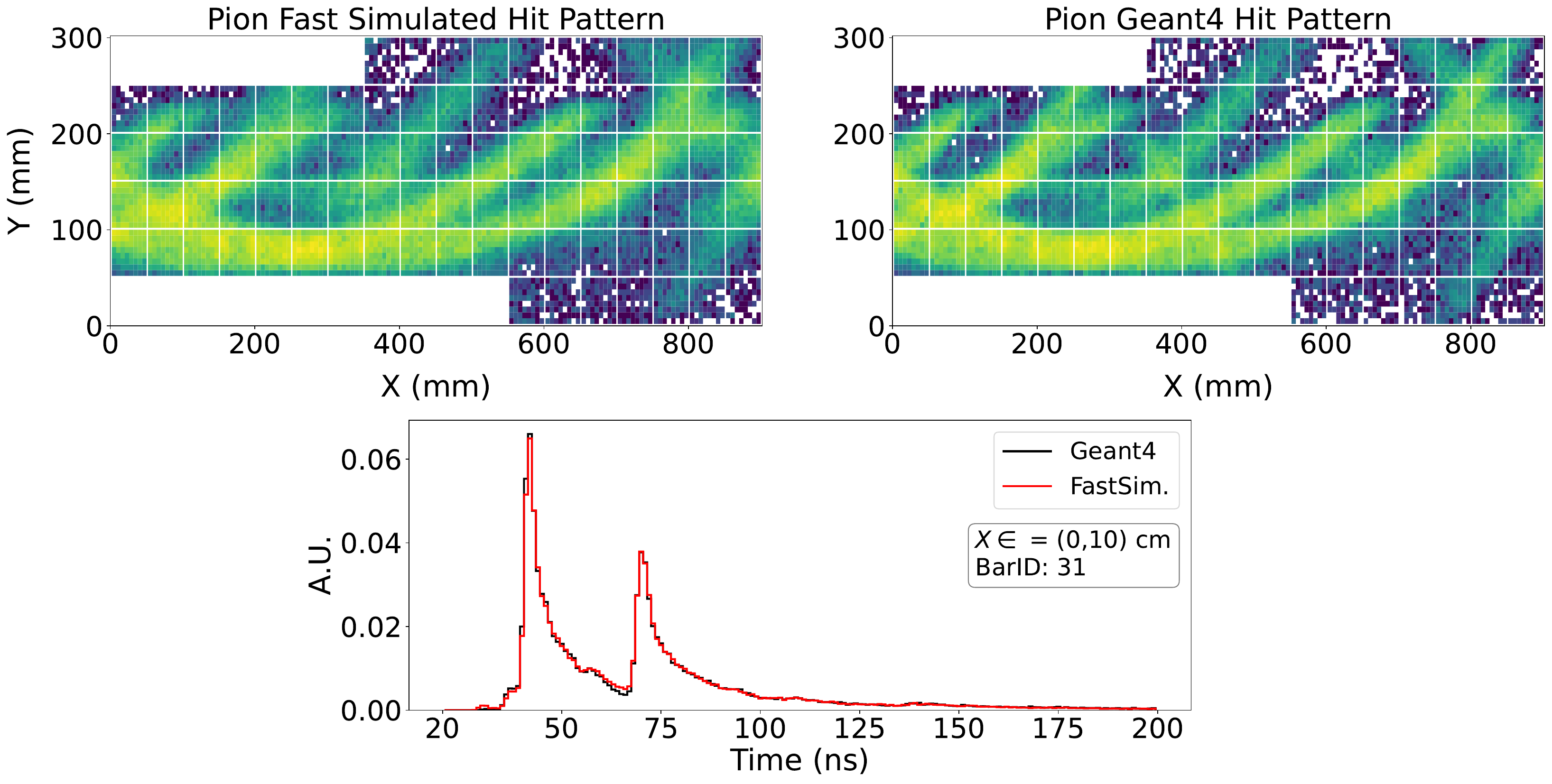}
    \includegraphics[width=0.49\textwidth]{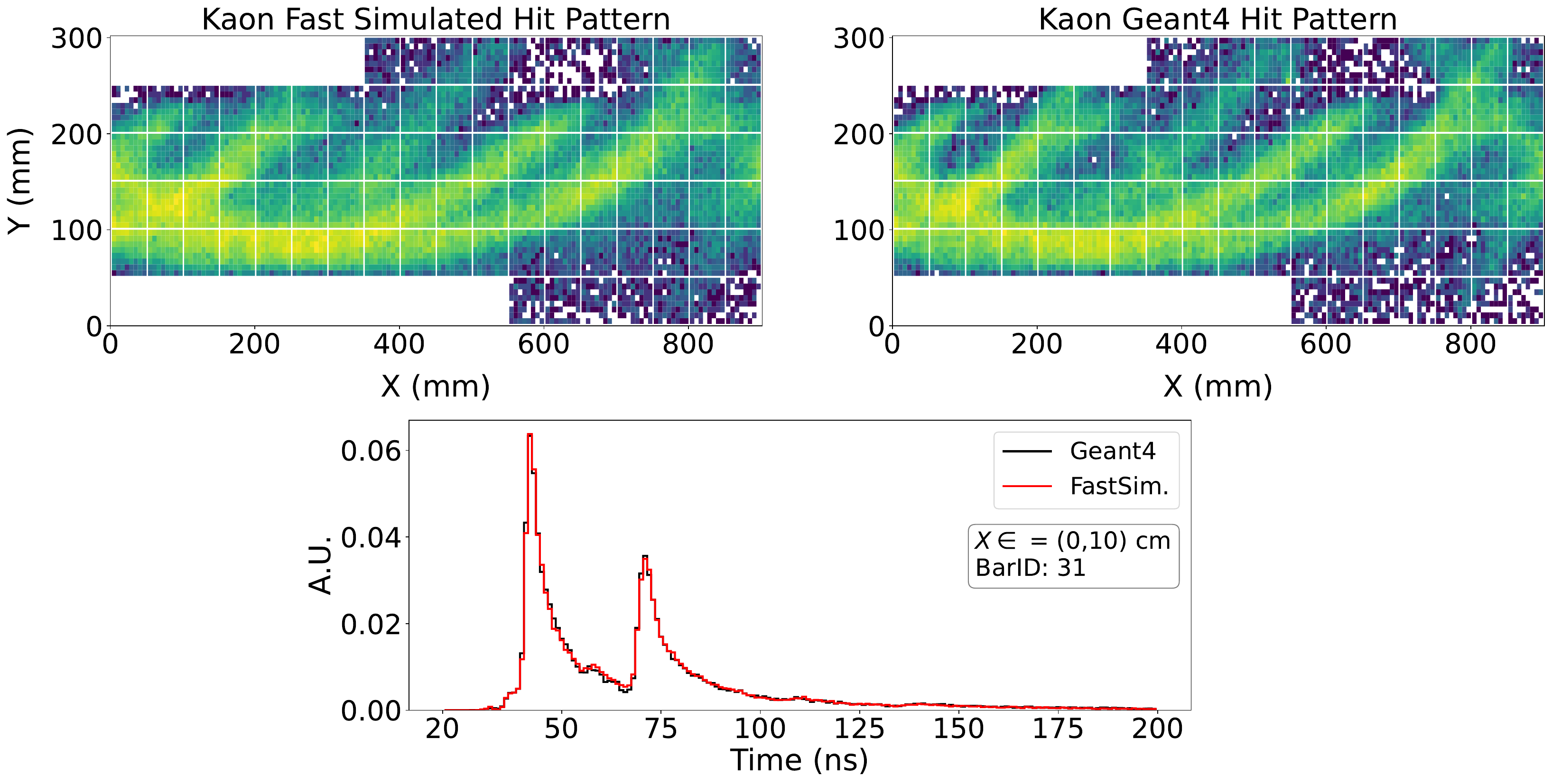}
    \caption{\textbf{Visual validation of generations}: Fast simulated hit patterns for pions (left column) and kaons (right column) at 
    $X \in (0\,\text{cm}, 10\,\text{cm})$ for bar 10 (top row) and bar 31 (bottom row), shown 
    alongside their Geant4 ground truth counterparts. The selected bars are centrally located in 
    the DIRC and span both optical boxes. Photon yield is consistent between fast-simulated and 
    ground truth samples in each case. The bottom panel shows the marginal time distribution for 
    a representative bar segment, with the fast simulation (red) closely reproducing the 
    characteristic double-peaked Cherenkov timing structure of the Geant4 reference (black).}
    \label{fig:generations}
\end{figure}

From visual inspection, the model faithfully reproduces the underlying spatial hit patterns and timing structure upon integration over multiple tracks. Additional generations at different bars and locations on bar faces can be found in \ref{app:add_gens}.

As qualitative validation of generation fidelity, we train the FM in classifier mode from scratch on both fast-simulated and \geant samples, following the procedure of~\cite{fanelli2024deep}.\footnote{Training from scratch refers to random weight initialization.} The model is then evaluated on the same \geant test set used throughout previous sections. Since the fast-simulated training sample is drawn from the learned probability distribution of the generative model, we augment it to twice the nominal training size in order to improve phase space coverage and partially mitigate statistical sparsity \cite{fanelli2024deep}.

Figure~\ref{fig:cls_gen_val} summarizes the use of classification as a probe of generation fidelity, comparing models trained on fast-simulated samples (red) and \geant samples (blue), alongside the geometrical reconstruction baseline (black), both integrated over the full phase space (left) and differentially in momentum (right).
\begin{figure}[!]
    \centering
    \includegraphics[width=0.49\textwidth]{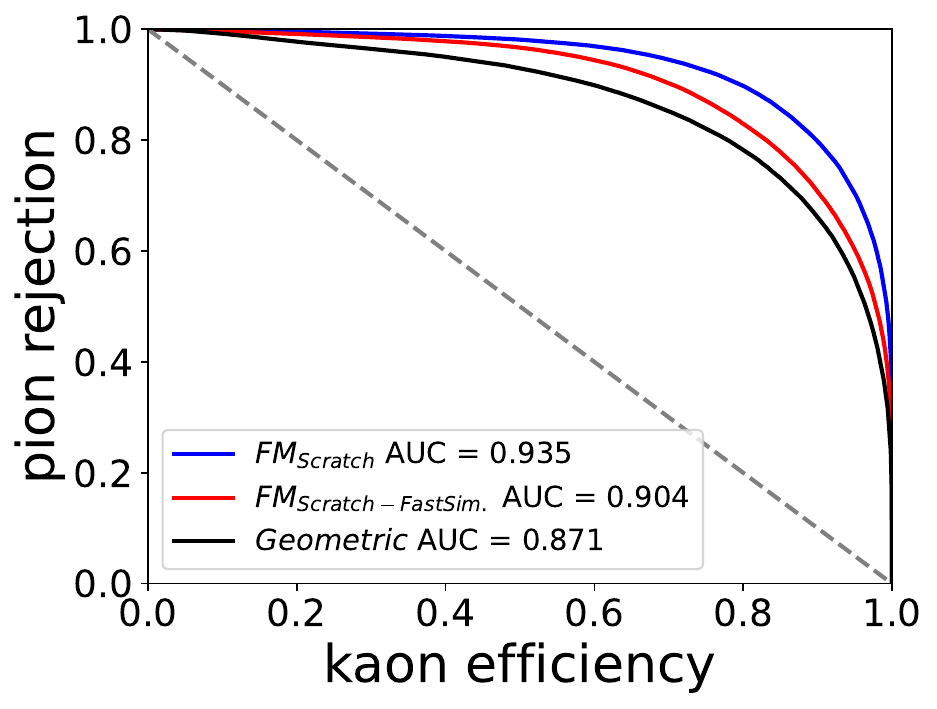}
    \includegraphics[width=0.48\textwidth]{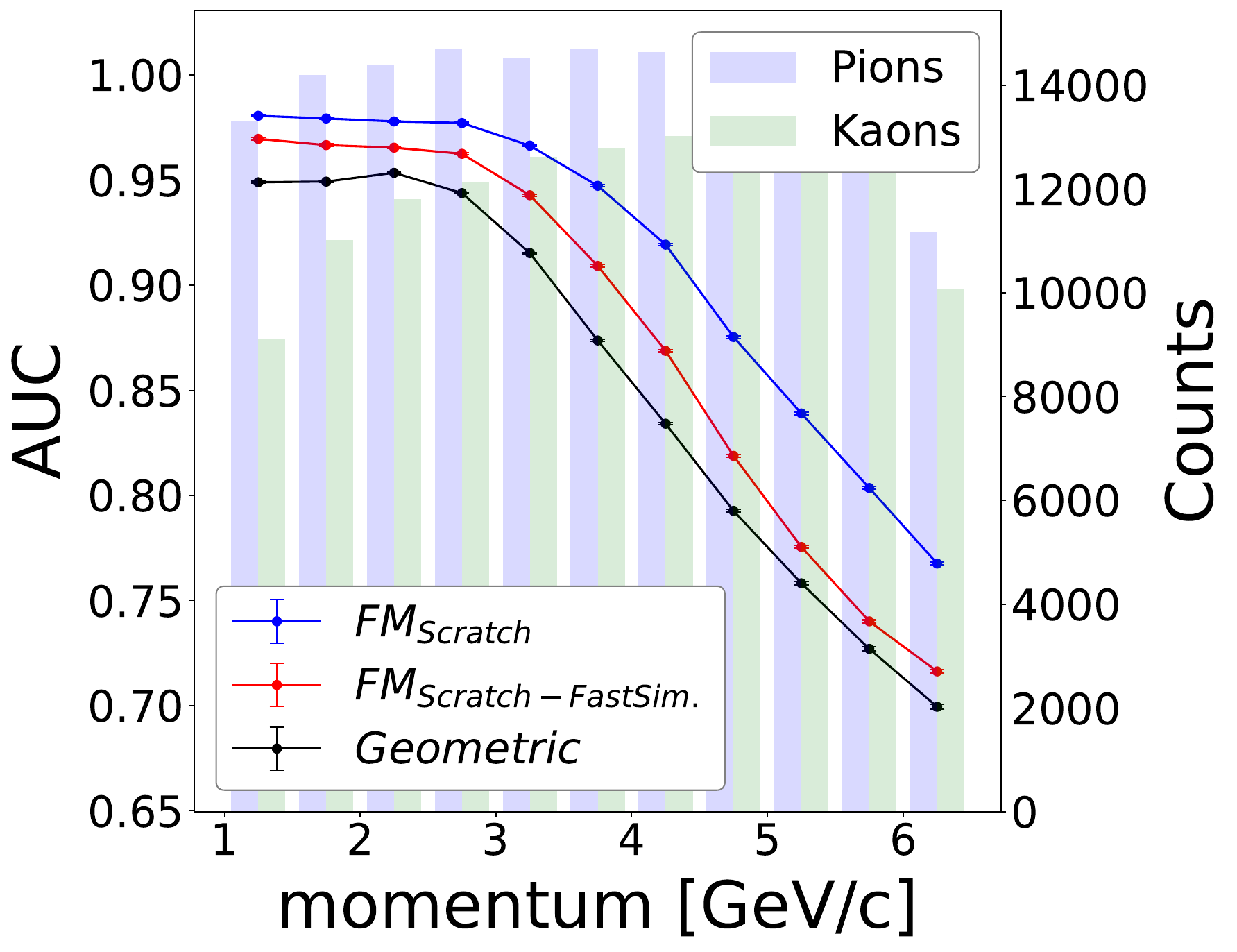}
    \caption{\textbf{Validation of generations via classification:} Pion rejection as a function 
    of kaon efficiency (left) for the foundation model trained on \geant samples 
    (blue), the foundation model trained on fast-simulated data 
    (red), and the geometrical reconstruction 
    baseline (black), integrated over the full phase space. AUC as a function of track momentum (right), 
    with uncertainties represented as 95\% quantiles obtained via bootstrapping. Pion and kaon 
    test counts are shown on the secondary axis.}
    \label{fig:cls_gen_val}
\end{figure}

When trained and evaluated on \geant, the classifier achieves an AUC of $0.935$, while the same procedure applied to fast-simulated data yields $0.904$, both exceeding the geometric reconstruction baseline of $0.871$. Interpreted in this context, the performance obtained using fast-simulated samples reflects the degree to which the generative model preserves class-discriminating structure present in the underlying detector response. The observed degradation therefore serves as a direct proxy for imperfections in the learned data distribution. Moreover, the results of the model trained from scratch inherently validate the fine tuning conclusions drawn in \cite{giroux2025towards}, as indicated by PID performance in the previous section where full fine tuning was employed.

When integrated over the full kinematic phase space, however, the overall deviation remains modest at the $\mathcal{O}(3\%)$ level, indicating that the model retains strong global fidelity despite localized degradation in the most challenging regions. Crucially, the observed behavior is consistent with a data-limited generative regime rather than a limitation of the classification procedure itself. The pre-training dataset spans the full acceptance of the \gluex DIRC but contains only $\sim 700$k samples per particle species. In the context of next-token prediction models, where performance typically scales strongly with both data volume and diversity \cite{kaplan2020scalinglawsneurallanguage}, this represents a relatively sparse sampling of a high-dimensional phase space.

As a consequence, certain regions of the phase space are underrepresented during training, leading to uneven coverage and an effective bias toward more frequently sampled track configurations. This imbalance manifests as a mild form of mode averaging or collapse in sparsely populated regimes, particularly at high momentum where subtle variations in Cherenkov ring structure are critical for particle separation. Importantly, this limitation is not fundamental to the modeling approach, but rather reflects the statistical constraints of the training corpus.
The momentum-dependent AUC reveals that the gap is predominantly driven by tracks above $\sim 3\,\text{GeV}/c$, where particle separation is intrinsically more difficult and relies on increasingly fine-grained features of the ring structures. In this regime, the generative model exhibits a progressive loss of discriminative detail, effectively blurring the distinction between pion and kaon signatures.

These results therefore suggest that the current performance should be interpreted as a lower bound on the achievable fidelity of the foundation model in this setting. Increasing the scale, uniformity, and diversity of the pre-training dataset is expected to systematically improve the resolution of fine-grained features across phase space, reducing the observed discrepancies and driving closer agreement with \geant.

%%%%%%%%%%% Now photon yield

Unlike methods proposed for DIRC detectors in~\cite{fanelli2024deep} or~\cite{giroux2025generativemodelsfastsimulation}, which require an auxiliary procedure to reproduce the photon yield, the foundation model demonstrated in this work directly learns the yield as part of the generative process~\cite{giroux2025towards}. Figure~\ref{fig:yields} shows the generated photon yield alongside the Geant4 ground truth across all bars of the \gluex DIRC for kaons (left) and pions (right).

\begin{figure}[!]
    \centering
    \includegraphics[width=0.5\textwidth]{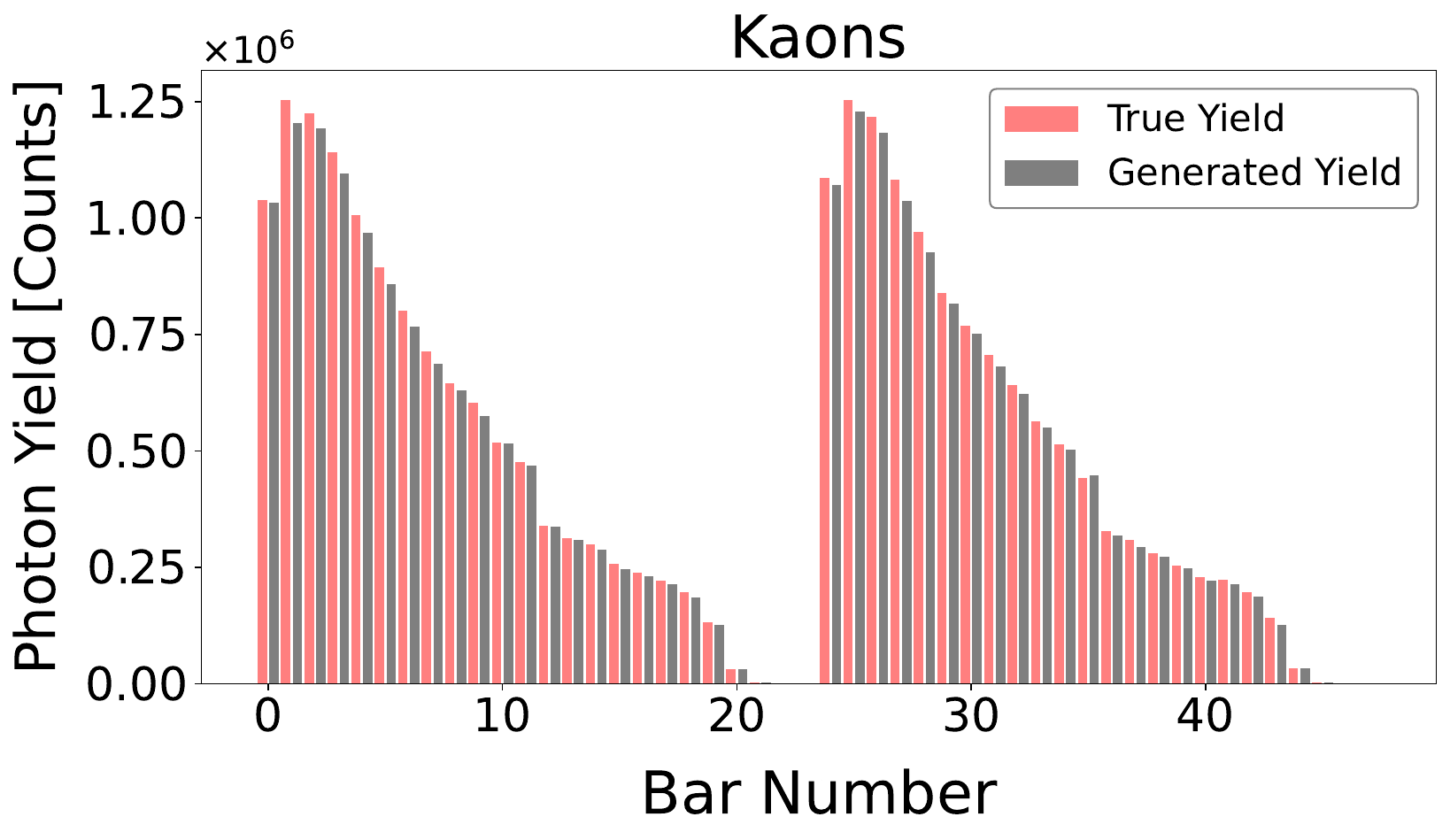}%
    \includegraphics[width=0.5\textwidth]{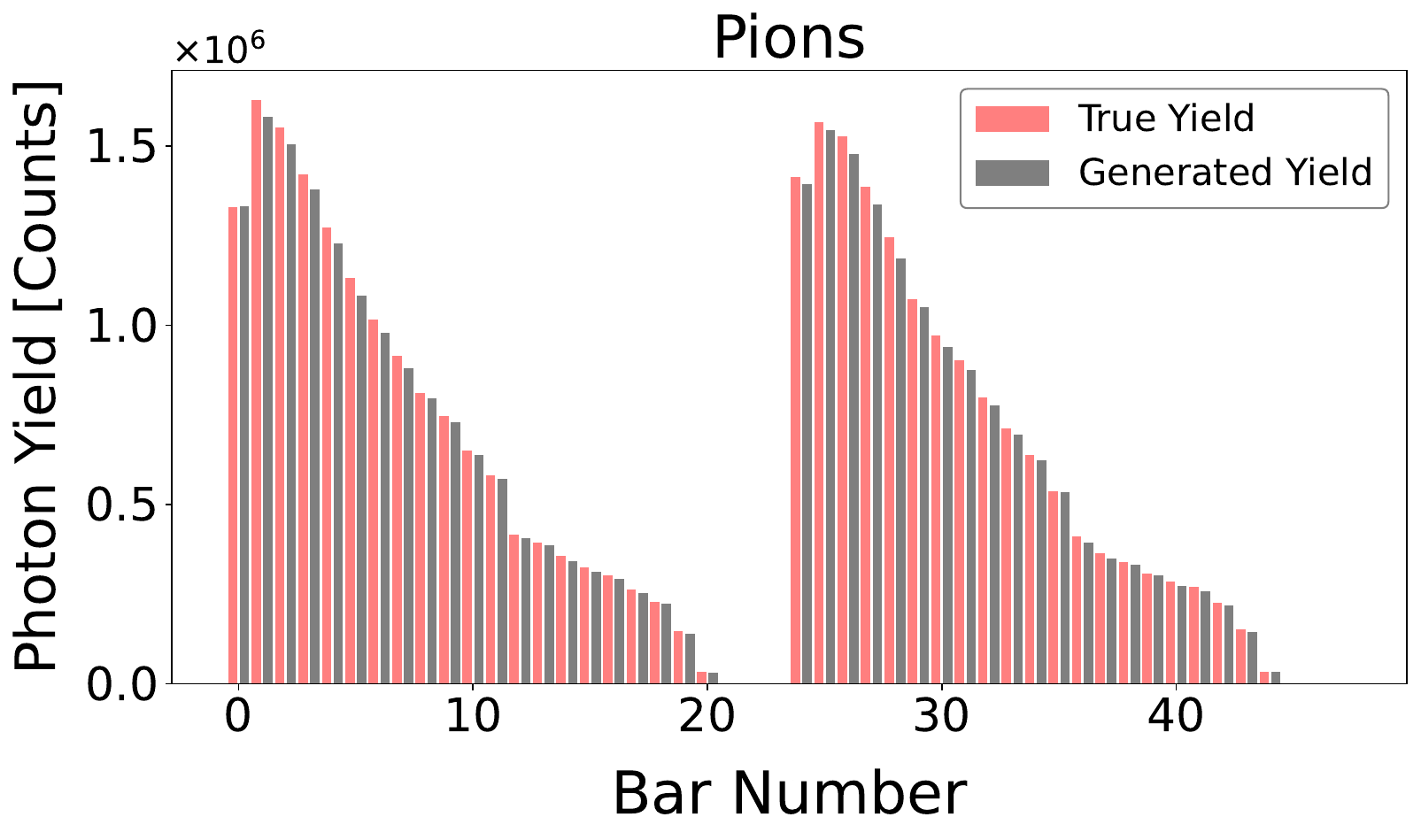}
    \caption{\textbf{Photon Yield:} Generated (black) and ground truth (red) photon yields as a function of bar number across the \gluex DIRC for kaons (left) and pions (right), integrated over the full phase space. The generated yield closely reproduces the ground 
    truth across all bars, with no auxiliary yield model required.}
    \label{fig:yields}
\end{figure}

From inspection of Figure~\ref{fig:yields}, the generated photon yield is in close agreement with the Geant4 ground truth across all bars, and for both particle species. The characteristic variation in yield across bar number, reflecting the geometric and optical properties of the detector, is faithfully reproduced by the foundation model. This is a particularly notable result, as the yield is learned implicitly through the autoregressive generation process rather than through any dedicated auxiliary network, representing a practical simplification over prior fast simulation approaches for DIRC detectors.

%%%%%%%%%%%%%%%%%%% Noise Filtering %%%%%%%%%%%%%%%%%%%%%%%%%%%%%%%%%%%
\subsection*{Noise Filtering}

Following the procedure outlined in~\cite{giroux2025towards}, we classify individual hits (token level) created by an incoming charged track as either signal or noise. The model is trained irrespective of class label, \textit{i.e.}, both pions and kaons are included under a single unified model. As noted in \cite{giroux2025towards}, full fine tuning from a pre-trained generative model provides no additional performance increase(in contrast to classification), and as such, we train from randomly intialized weights.

We use a simulated dark rate of $\SI{25}{\kilo\hertz}/\text{cm}^2$ in the \gluex DIRC PMTs, sampled randomly during training. For a given track, both the ordering in time and the number of dark photons are stochastic, requiring the architecture to learn contextual representations of signal and noise rather than relying on positional information. During evaluation, we follow the same sampling procedure, generating 50 unique representations of each track in the test set to obtain uncertainty estimates. Performance is evaluated integrated over the entire phase space, in contrast to the fixed-momentum evaluation of~\cite{giroux2025towards}. 

Figure~\ref{fig:Filtering} shows the performance of our method for kaons and pions in terms of precision-recall curves and noise rejection as a function of signal retention, with 99\% confidence intervals obtained via bootstrapping. The precision-recall curves are insensitive to the inherent class imbalance.

\begin{figure}[!]
    \centering
    \begin{subfigure}[b]{0.49\textwidth}
        \includegraphics[width=0.49\textwidth]{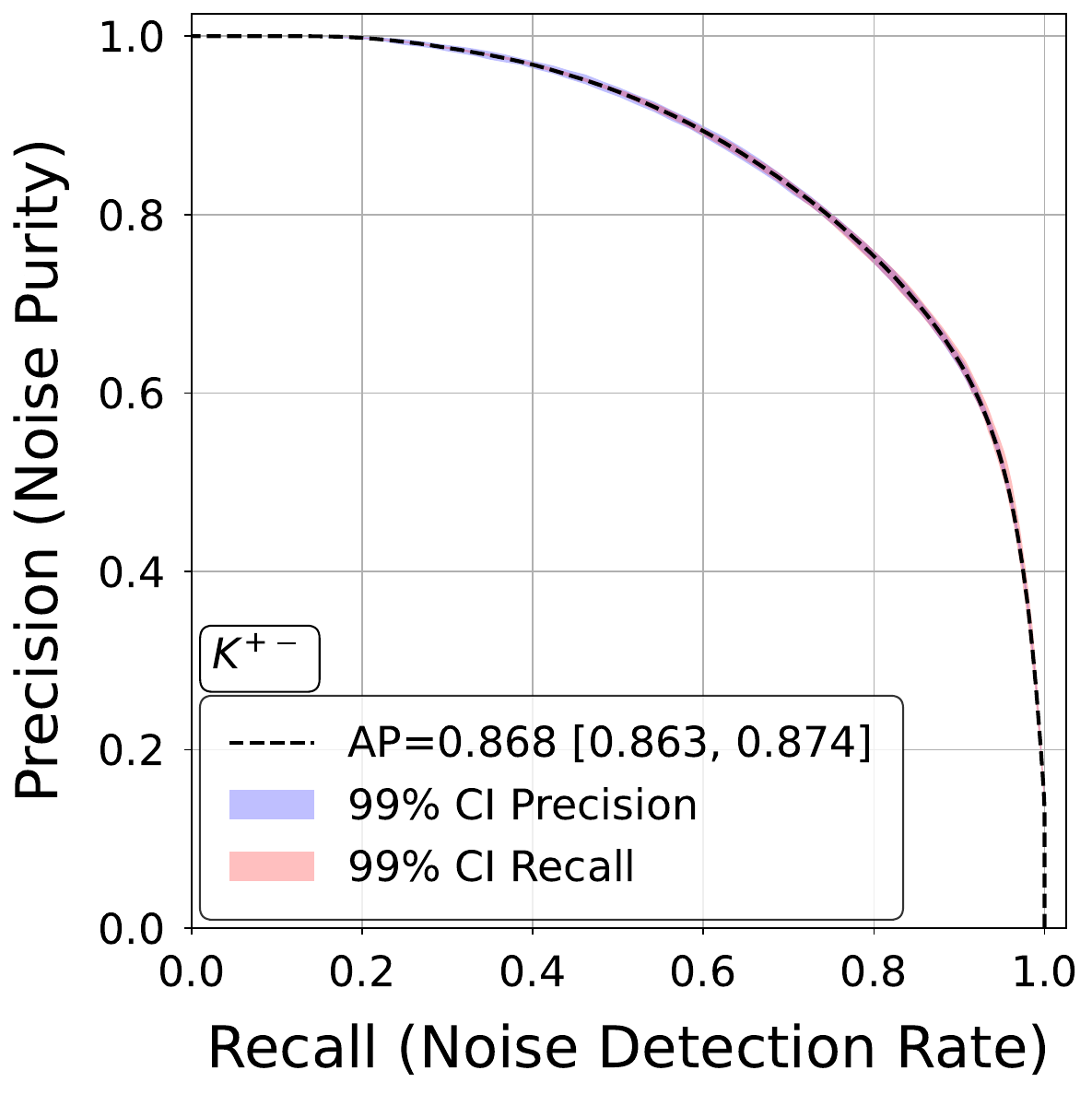}
        \includegraphics[width=0.49\textwidth]{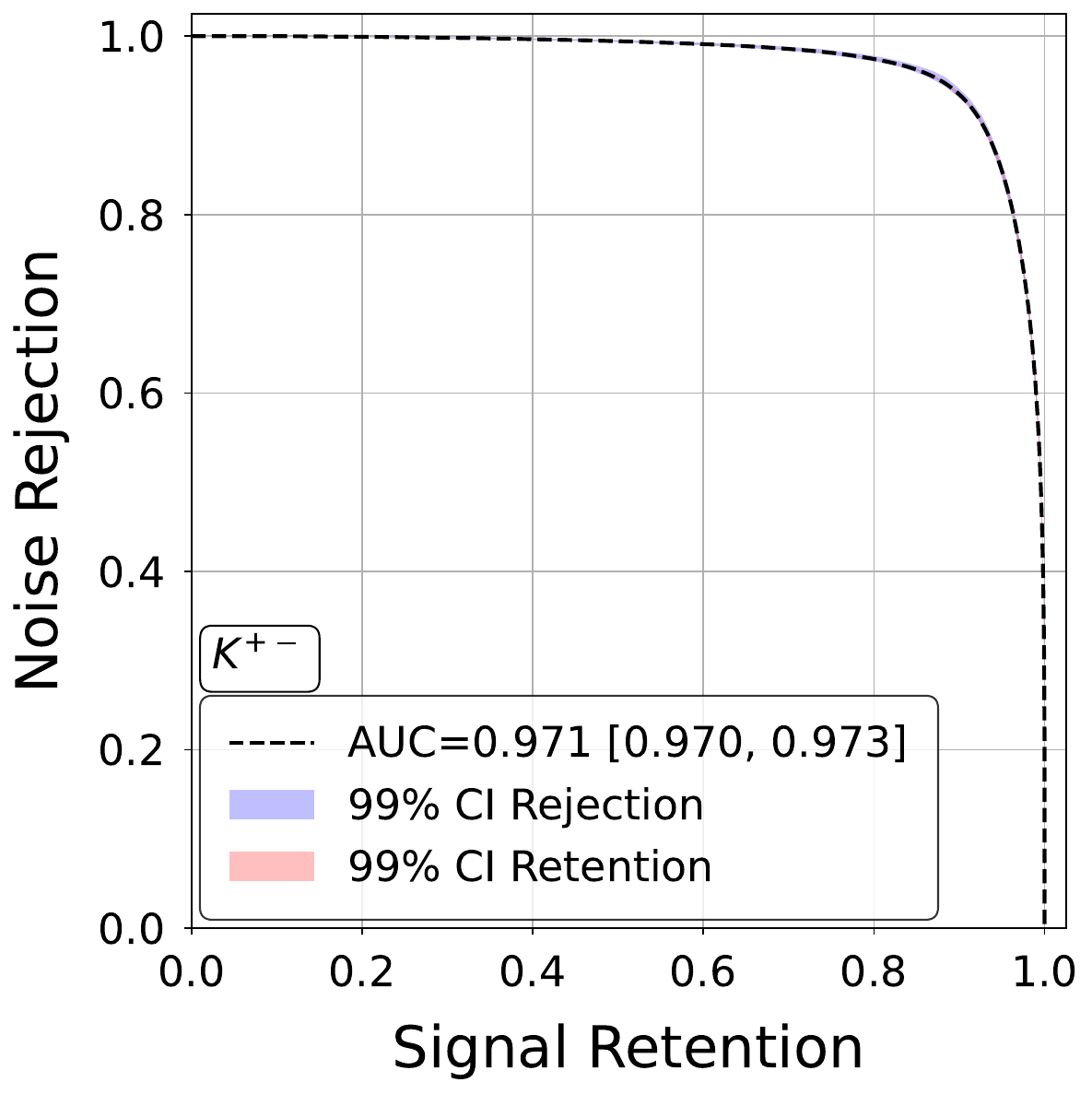}
        \caption{Kaons}
    \end{subfigure}
    \begin{subfigure}[b]{0.49\textwidth}
        \includegraphics[width=0.49\textwidth]{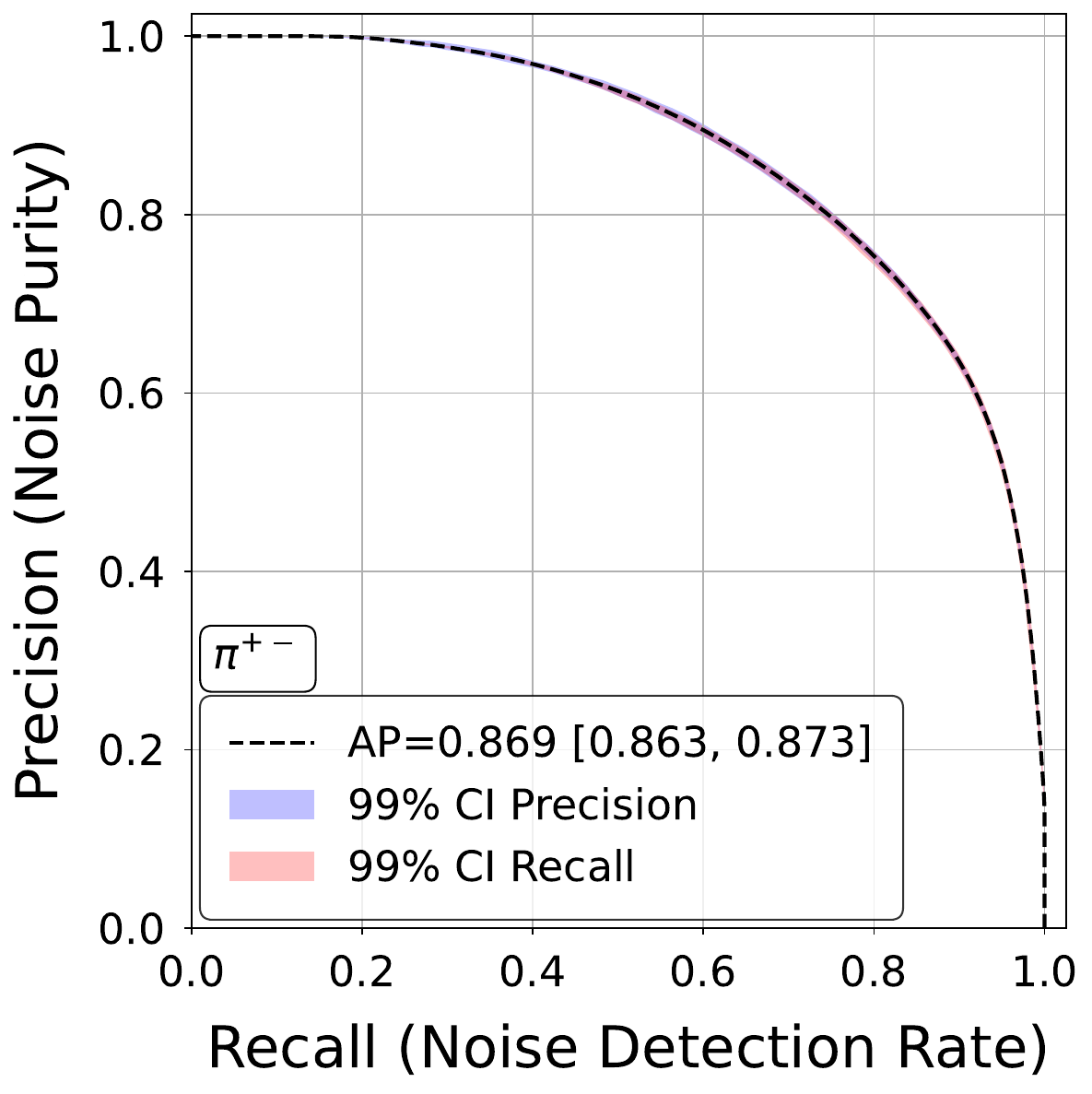}
        \includegraphics[width=0.49\textwidth]{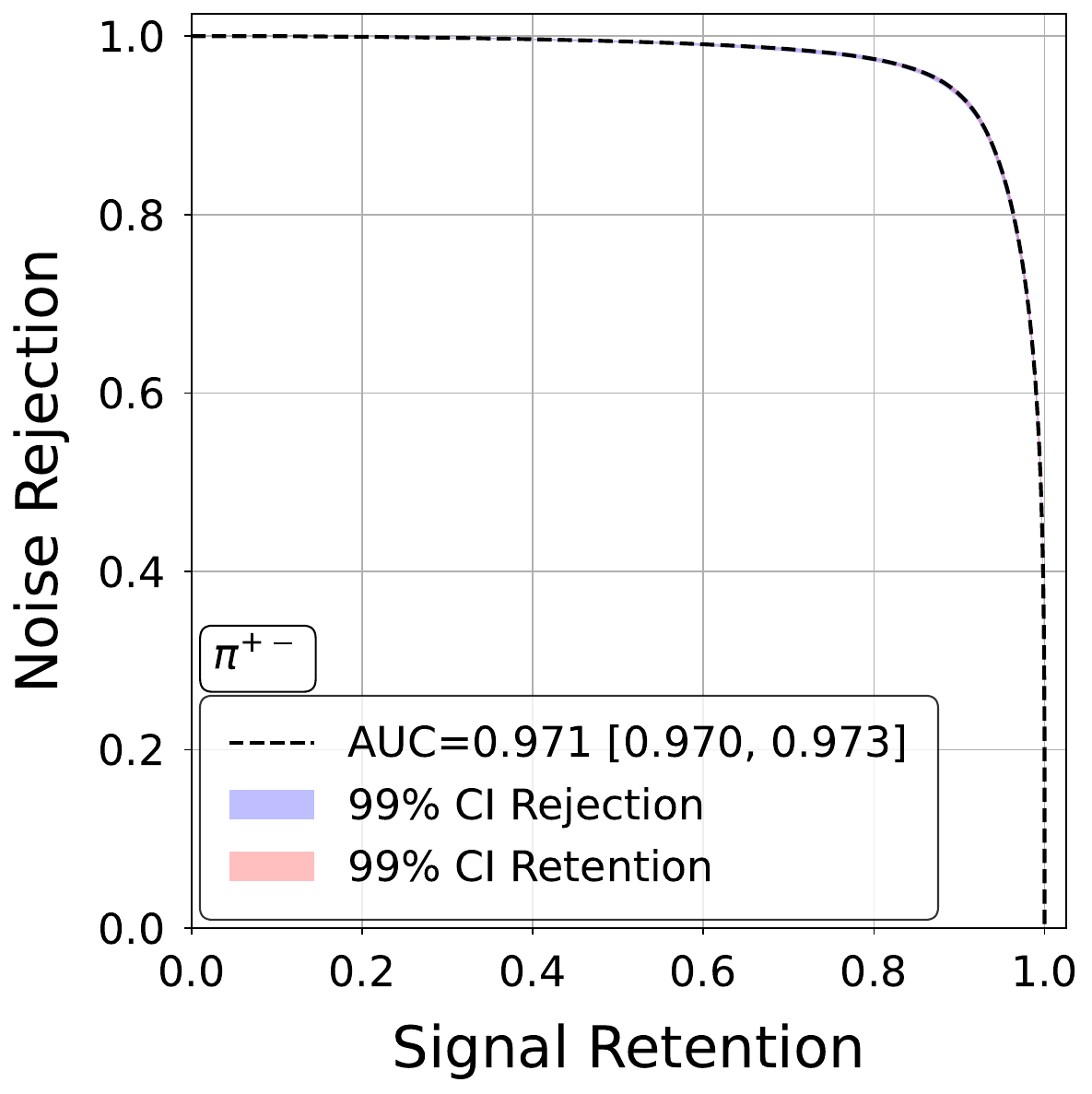}
        \caption{Pions}
    \end{subfigure}
    \caption{\textbf{Noise Filtering:} Noise filtering performance 
    integrated over the entire phase space for (a) kaons and (b) pions, shown as 
    precision-recall curves and noise rejection as a function of signal retention. The 
    precision-recall curves are insensitive to class imbalance. Shaded bands denote 99\% 
    confidence intervals.}
    \label{fig:Filtering}
\end{figure}

From inspection of Figure~\ref{fig:Filtering}, the model effectively discriminates detector noise from signal at the token level across the full phase space. The precision-recall curves yield an average precision (AP) of $\sim 0.869$ for pions and $\sim 0.868$ for kaons, with tight 99\% confidence intervals in both cases, indicating robust and stable performance. The noise rejection curves achieve an AUC of $0.971$ for both particle species, reflecting near-ideal signal retention at high noise suppression — a notably strong result given that the evaluation is integrated over all momenta. The reduced dark rate of $\SI{25}{\kilo\hertz}/\text{cm}^2$ relative to the $\SI{100}{\kilo\hertz}/\text{cm}^2$ used in~\cite{giroux2025towards} reflects the operating conditions of the \gluex DIRC, yet the consistency between pion and kaon performance confirms that the model generalizes well across particle types within a single unified network. 

%% file: 6_summary.tex
\section{Summary and Conclusions}\label{sec:summary}

In this work, we have demonstrated that a Mixture-of-Experts-based foundation model provides a unified and effective framework for the analysis of GlueX DIRC data, encompassing fast simulation, particle identification, and hit-level noise filtering within a single shared architecture. By operating directly on low-level detector inputs and leveraging a common transformer backbone across tasks, the approach eliminates the need for task-specific models while exceeding the performance of established methods.

A central feature of this framework is the integration of generative pre-training with fully fine-tuned downstream tasks. The model is first trained autoregressively to learn the underlying detector response, and the resulting weights are then used to initialize subsequent task-specific models. This pre-training stage provides a strong, physically grounded initialization of the latent space, leading to consistent performance gains of approximately $2\%$ in AUC for particle identification relative to training from scratch. In this sense, the generative objective serves not only as a tool for fast simulation, but as a mechanism for learning transferable representations that benefit downstream tasks.

The MoE architecture further enables class-conditional specialization at the generative level, while preserving shared representations that transfer efficiently across tasks. Together, these design choices allow the model to learn a coherent, event-level representation of the detector response, rather than fragmenting the problem into independent pipelines.

Across all evaluated tasks, the foundation model exhibits strong and consistent performance. For particle identification, the model achieves an AUC of $0.952$, outperforming both prior deep learning approaches and the geometrical reconstruction baseline, with the largest gains observed in the high-momentum regime where traditional methods degrade. In the context of fast simulation, the model produces high-fidelity Cherenkov photon patterns, accurately reproducing both the spatial structure and temporal characteristics of the detector response. Notably, the photon yield is learned implicitly as part of the autoregressive generation process, removing the need for auxiliary modeling and simplifying the simulation pipeline. For noise filtering, the model achieves near-ideal discrimination between signal and detector noise at the hit level, with an AUC of $0.971$ and stable performance across particle species.

To further probe the quality of the learned generative distribution, we provided a classification-based validation procedure, in which models trained on fast-simulated samples are evaluated against \geant ground truth. Within this framework, the performance gap between classifiers trained on generated versus true data provides a direct measure of the fidelity of the underlying simulations. We observe a modest but systematic degradation $\mathcal{O}(3\%)$, concentrated at higher momenta where the detector response is most sensitive to fine-grained structure. This behavior indicates that the current limitations of the generative model are not architectural, but instead reflect incomplete statistical coverage of the underlying phase space. The results obtained can therefore be viewed as a lower bound on generation fidelity given the scaling ability of auto-regressive architectures.

Taken together, these results establish the foundation model as both a high-performance and highly flexible alternative to traditional DIRC analysis pipelines. At the same time, they highlight a key scaling direction: the fidelity of the generative component — and, by extension, the unified framework as a whole — is presently constrained by the scope of the pre-training data. As larger and more diverse datasets become available, the model is expected to more accurately capture rare and high-complexity configurations, leading to convergence between fast-simulated and \geant-level performance without requiring modification to the architecture.

More broadly, this work reinforces the emerging paradigm of foundation models in experimental physics: that a single, well-trained model can serve as a general-purpose representation of detector data, capable of supporting a wide range of downstream tasks through full, or in certain cases partial fine-tuning. In the context of the \gluex DIRC, this paradigm offers a practical path toward scalable, maintainable, and high-fidelity analysis workflows, with clear potential for extension to other detector systems.

%% file: Appendix.tex
%\newpage
\appendix

\section{Additional Generations}\label{app:add_gens}

\begin{figure}[h]
    \centering
    \includegraphics[width=0.49\textwidth]{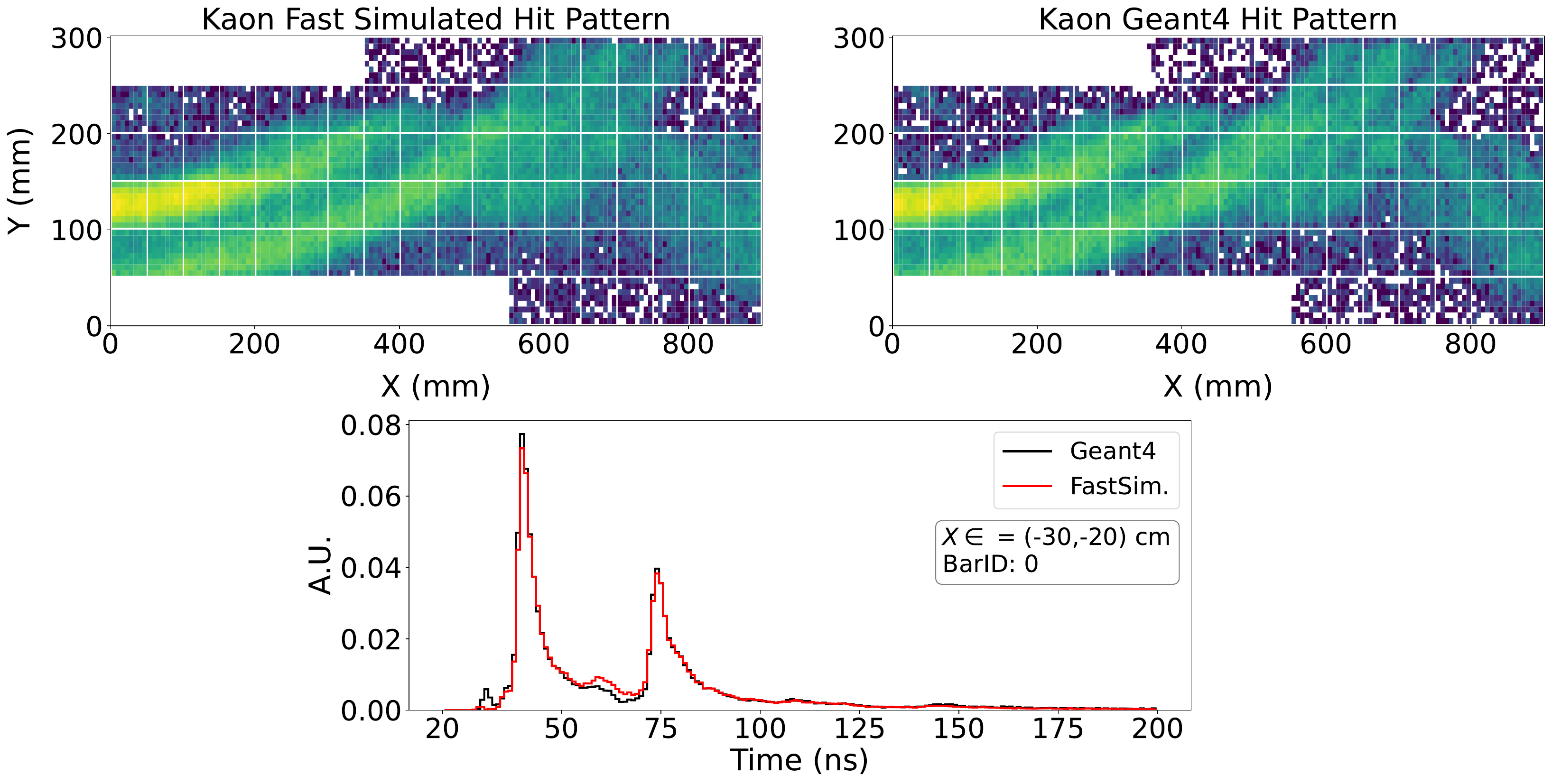}
    \includegraphics[width=0.49\textwidth]{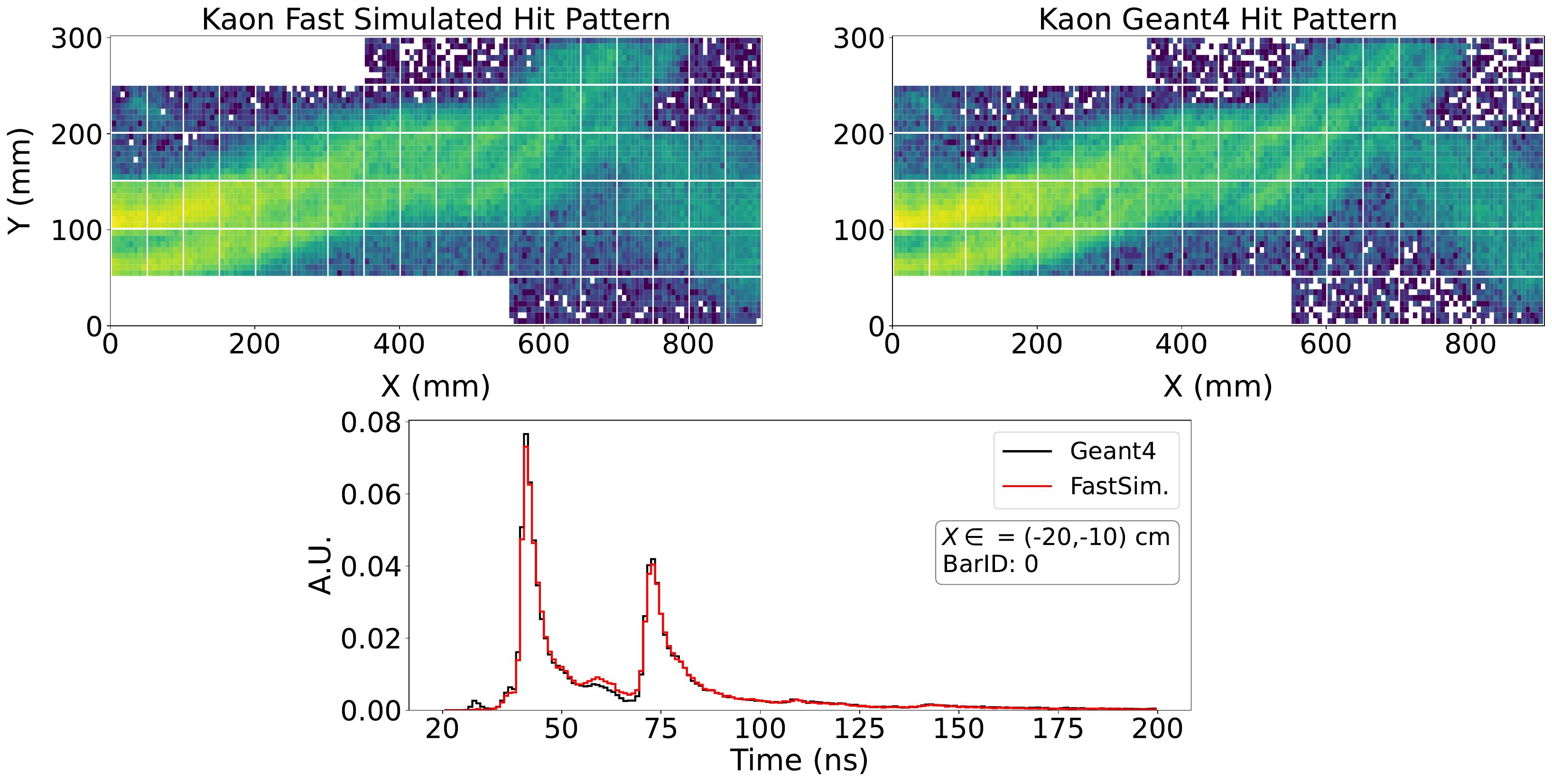}
    \includegraphics[width=0.49\textwidth]{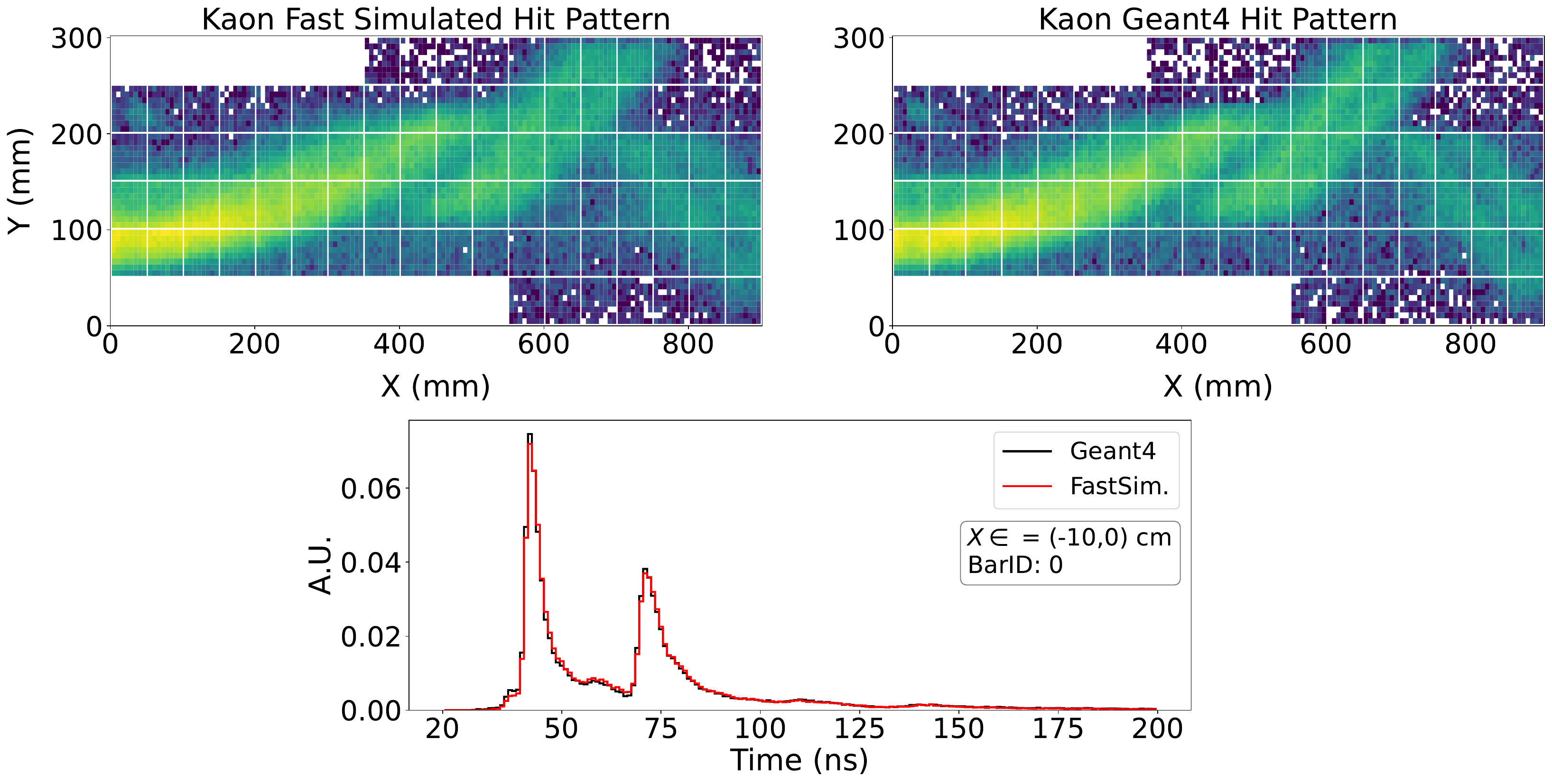}
    \includegraphics[width=0.49\textwidth]{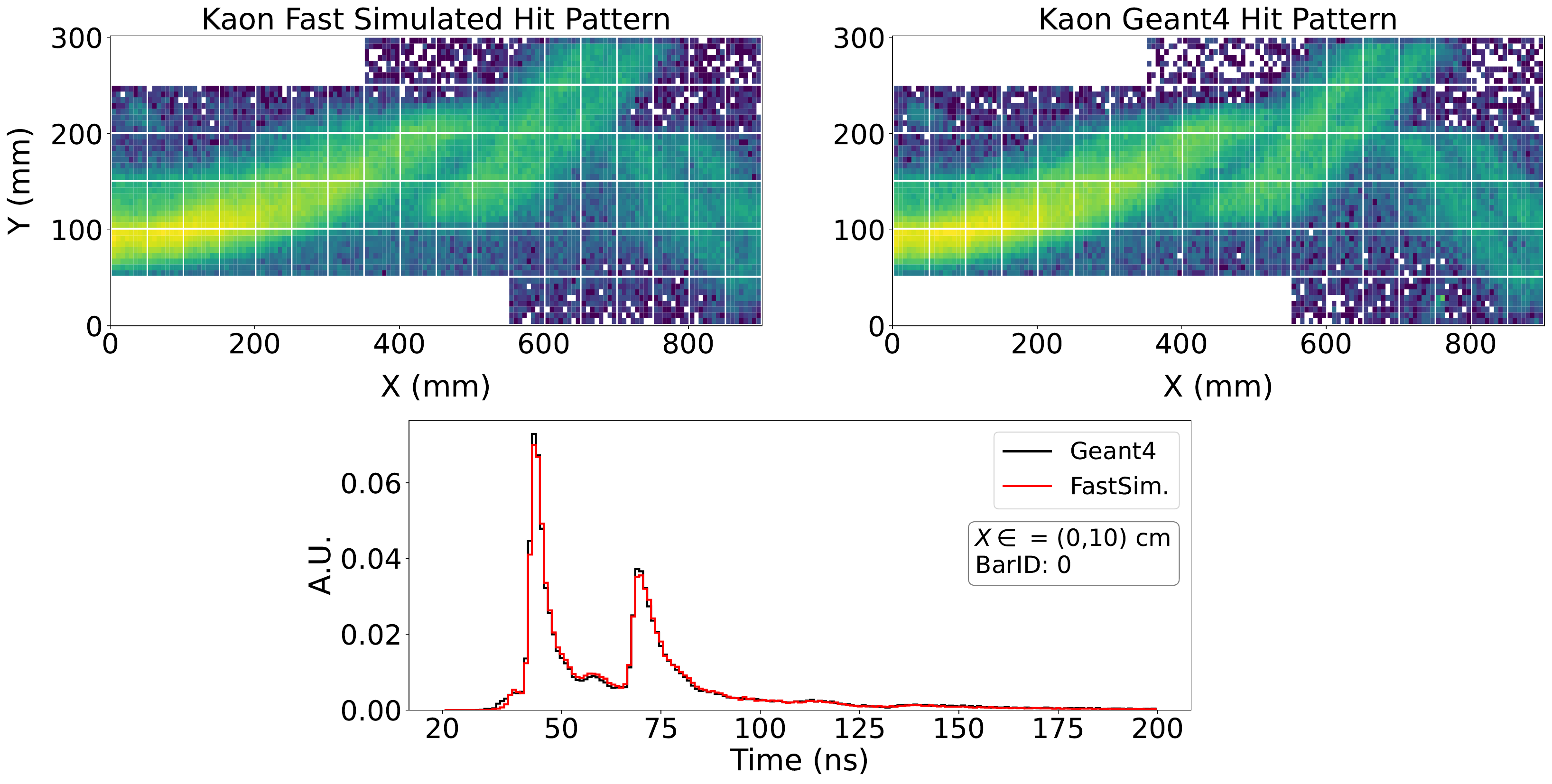}
    \includegraphics[width=0.49\textwidth]{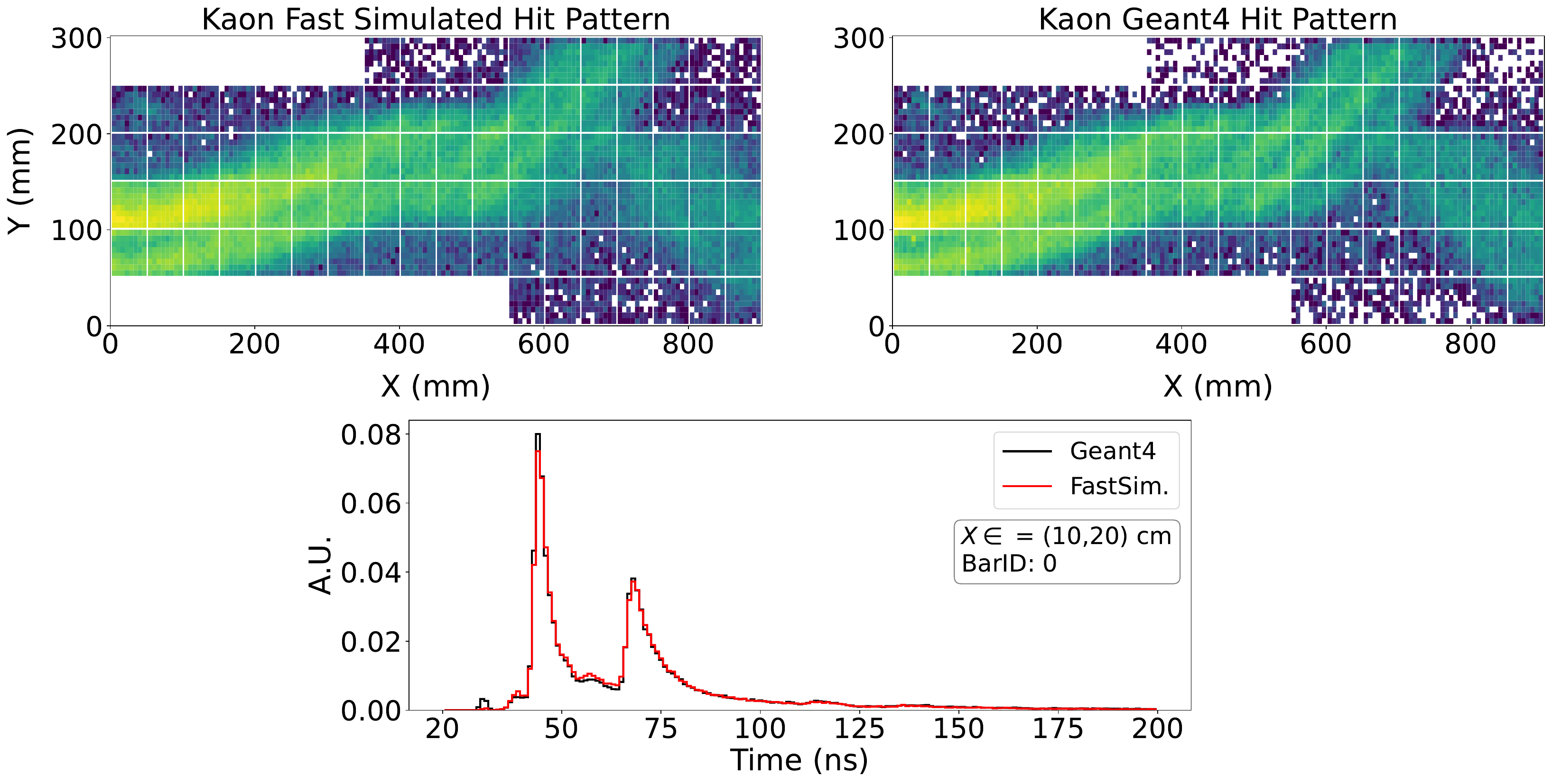}
    \includegraphics[width=0.49\textwidth]{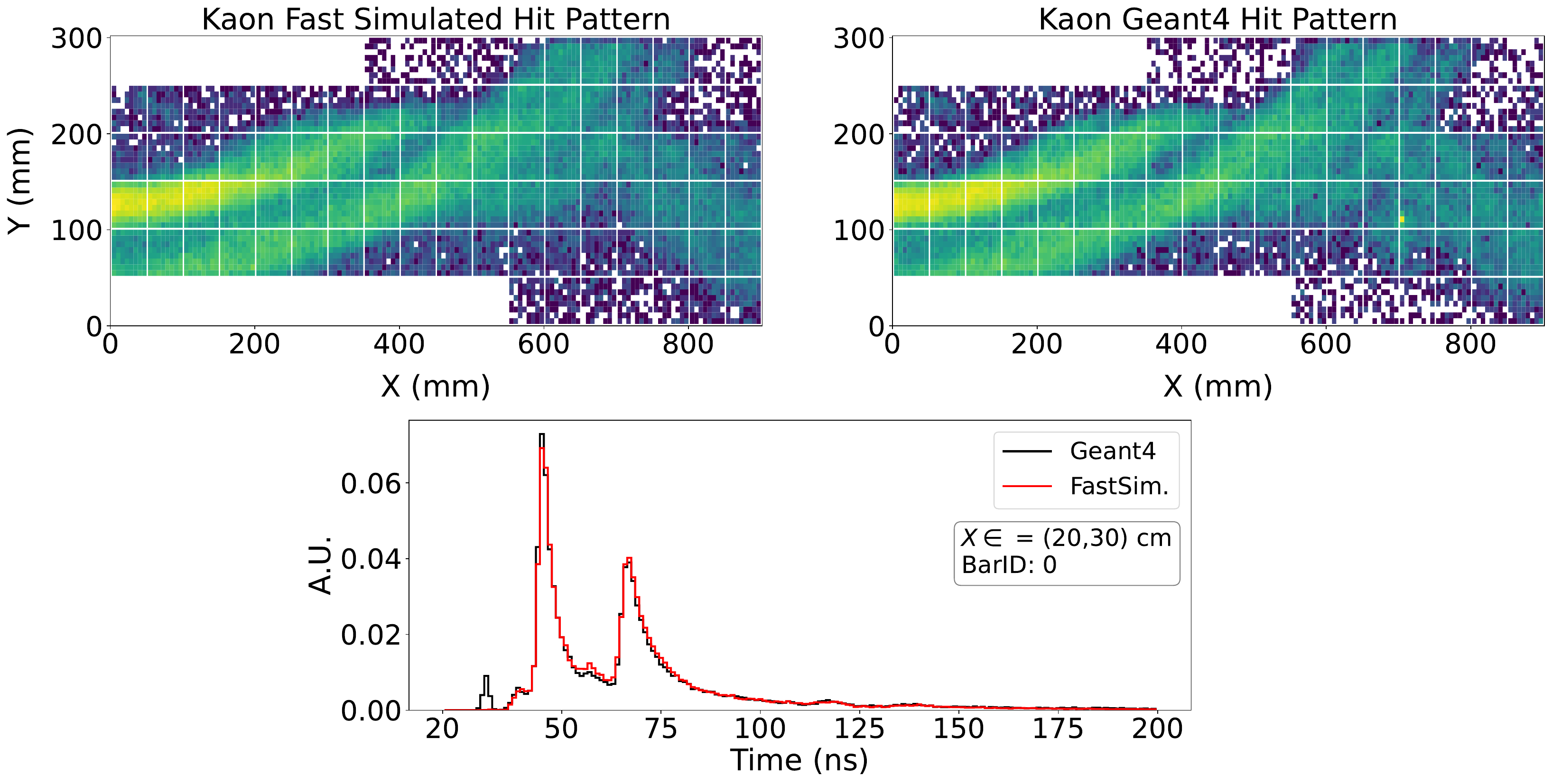}
    \caption{\textbf{Kaon generations at bar 0}: Fast-simulated and \geant hit patterns shown across all positions along the bar face, scanned sequentially from left to right, top to bottom.}
    \label{fig:app_kaon_bar0}
\end{figure}

\begin{figure}[h]
    \centering
    \includegraphics[width=0.49\textwidth]{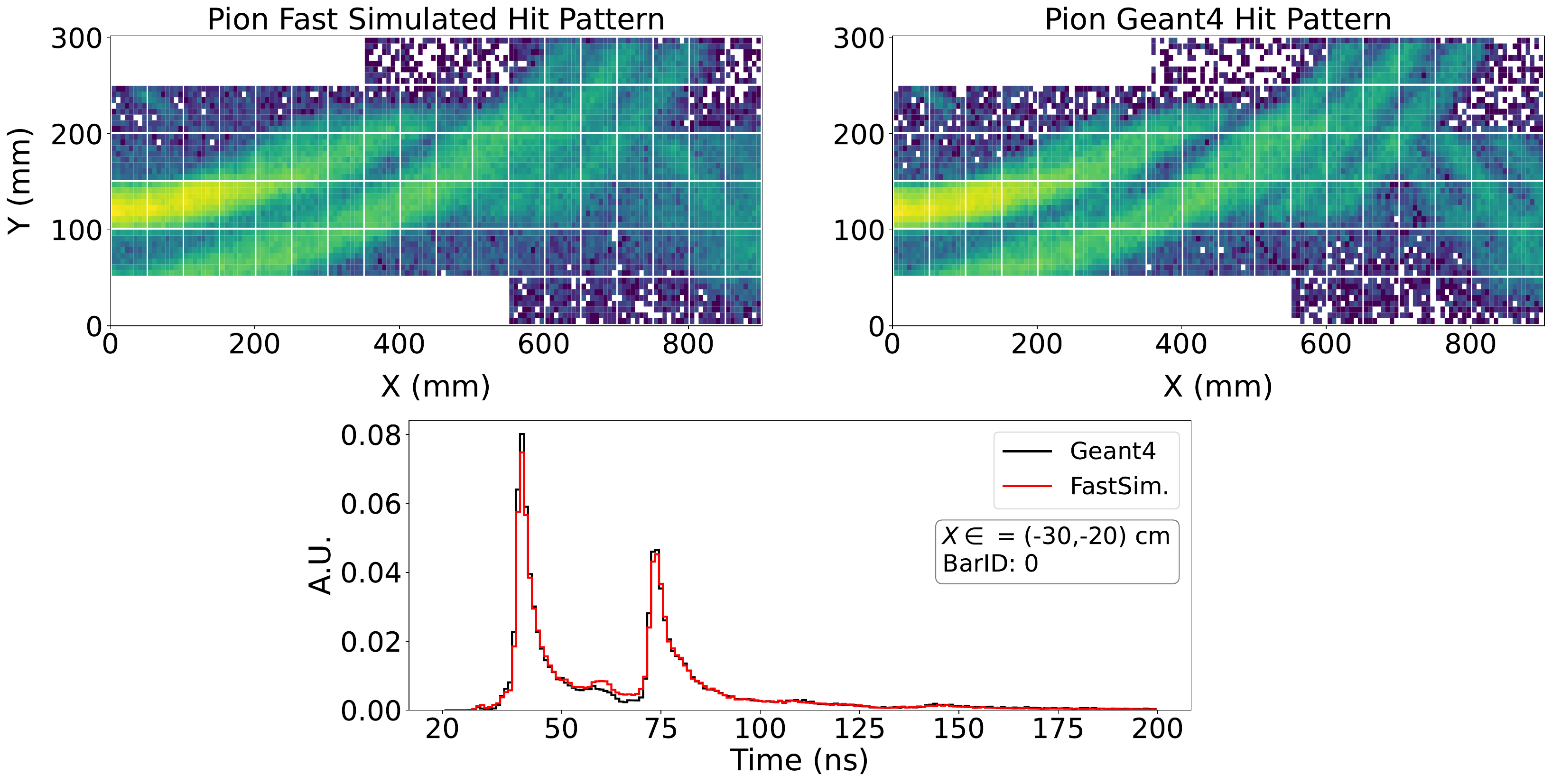}
    \includegraphics[width=0.49\textwidth]{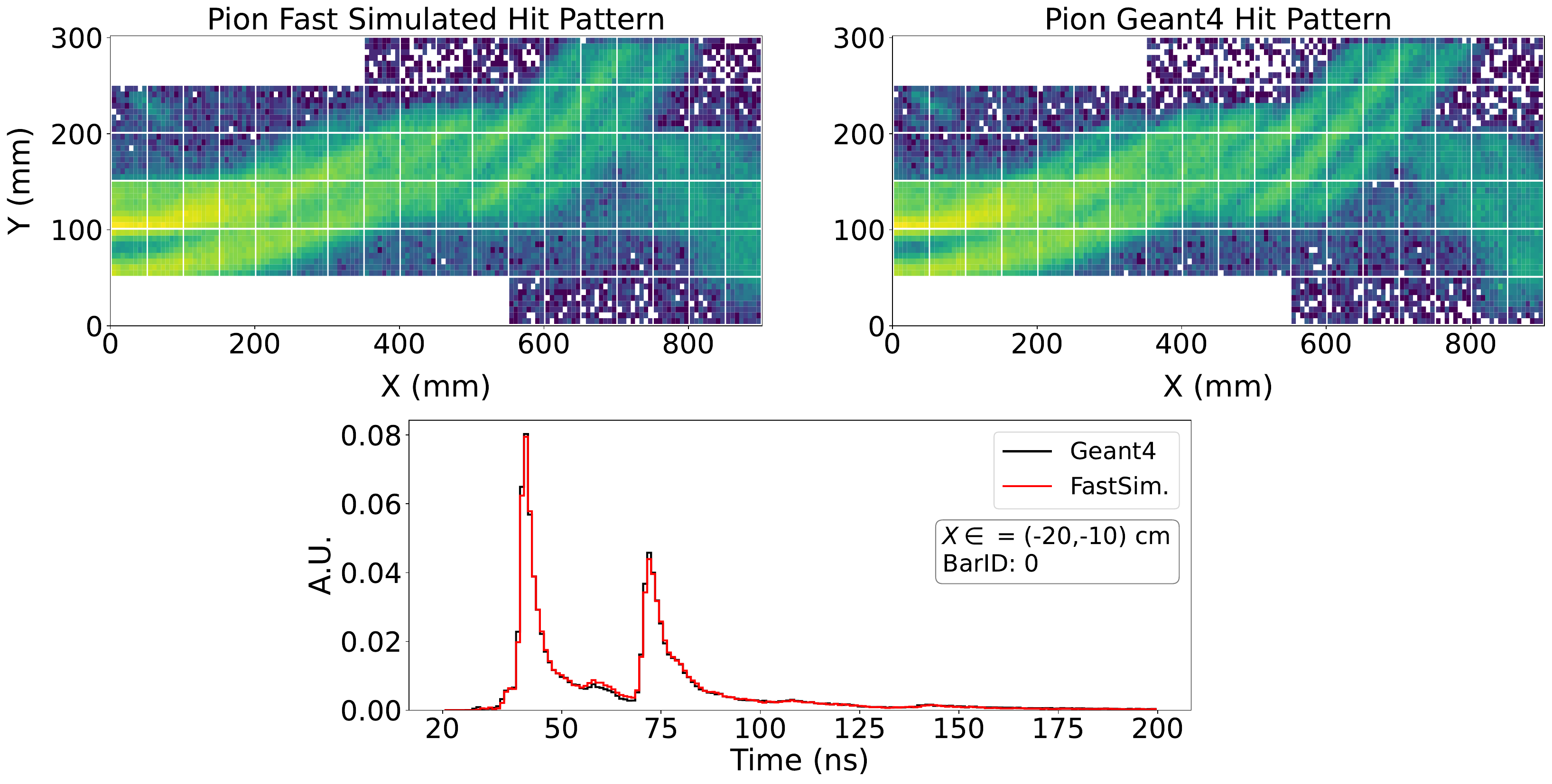}
    \includegraphics[width=0.49\textwidth]{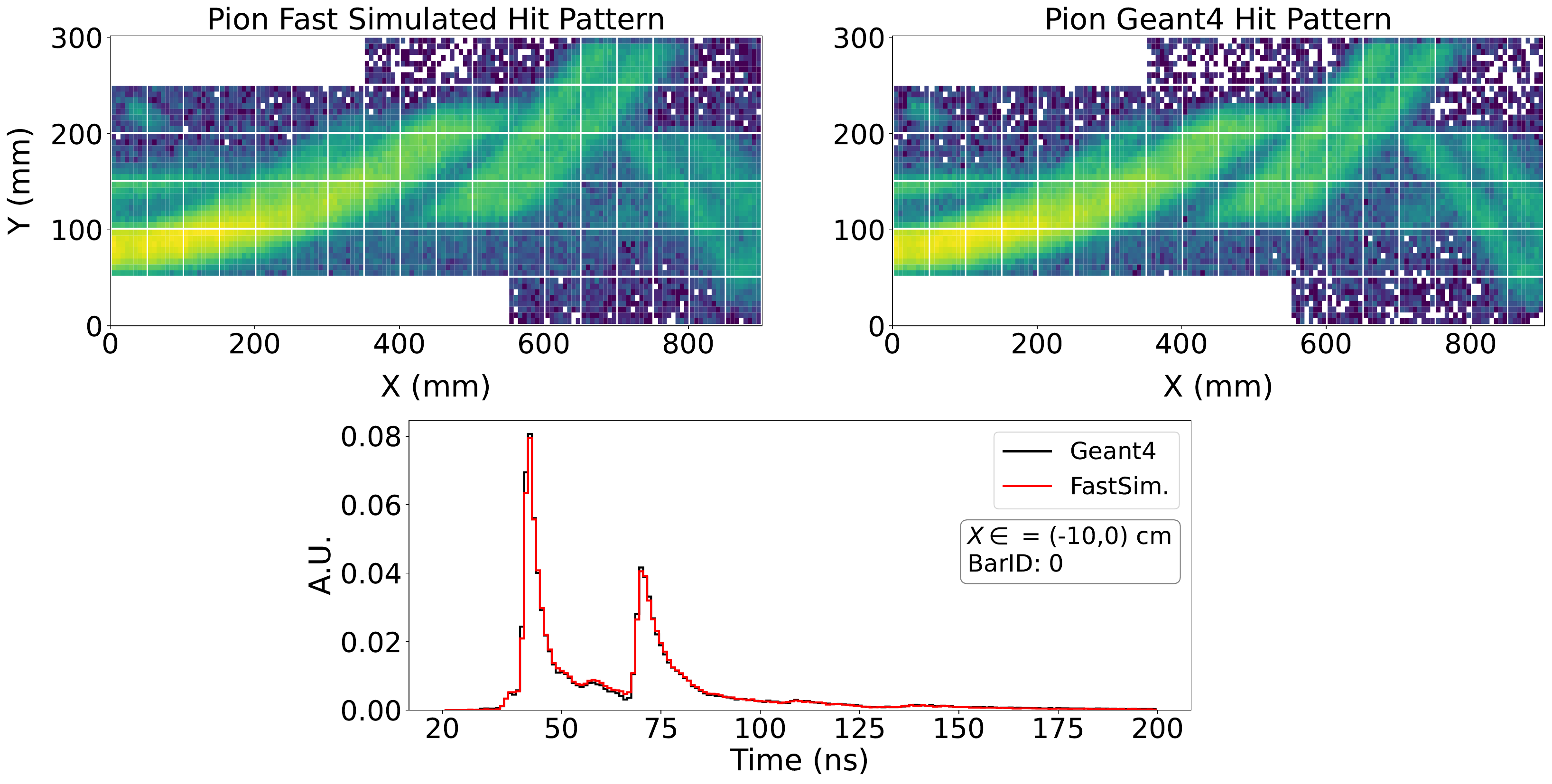}
    \includegraphics[width=0.49\textwidth]{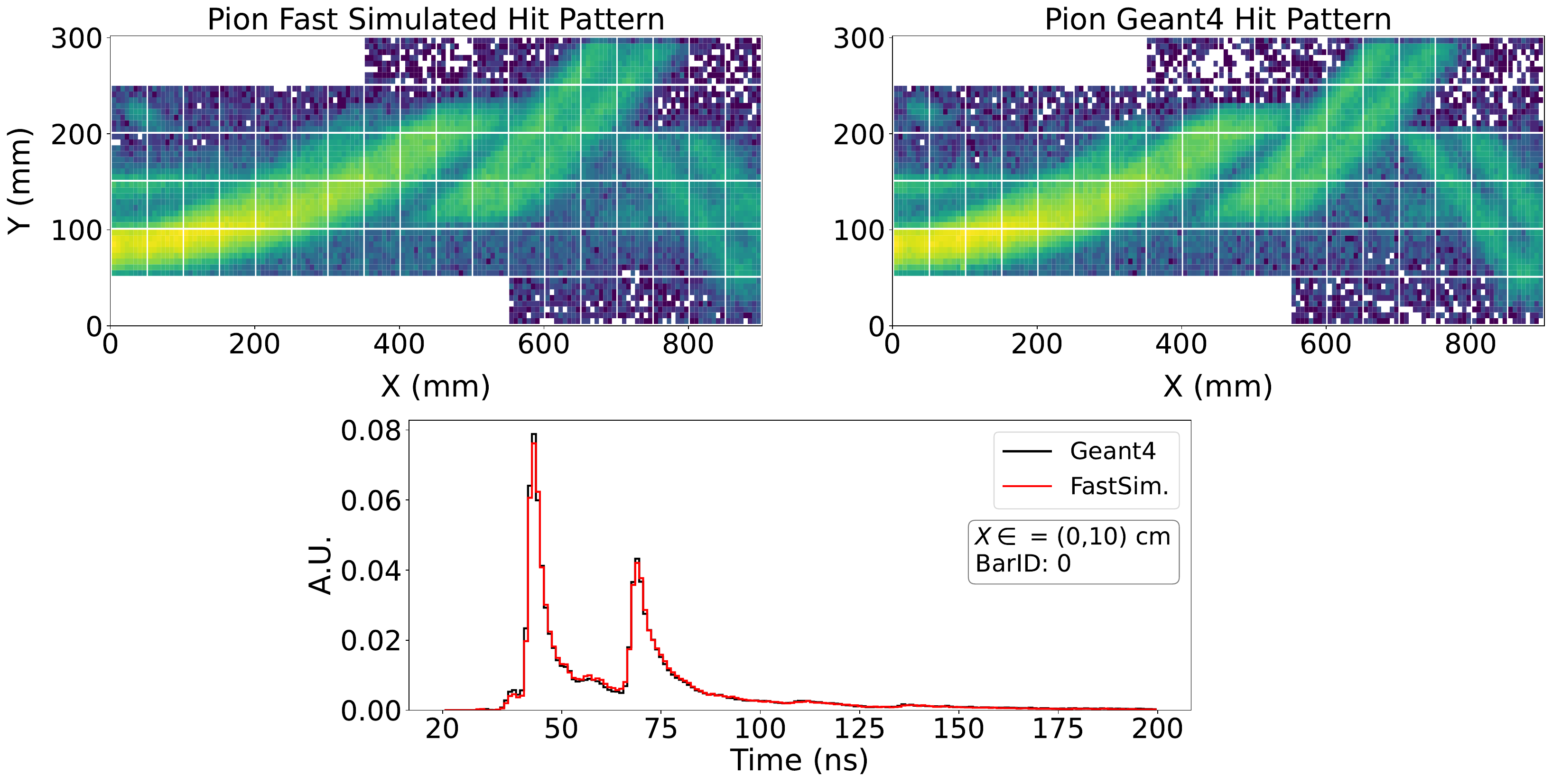}
    \includegraphics[width=0.49\textwidth]{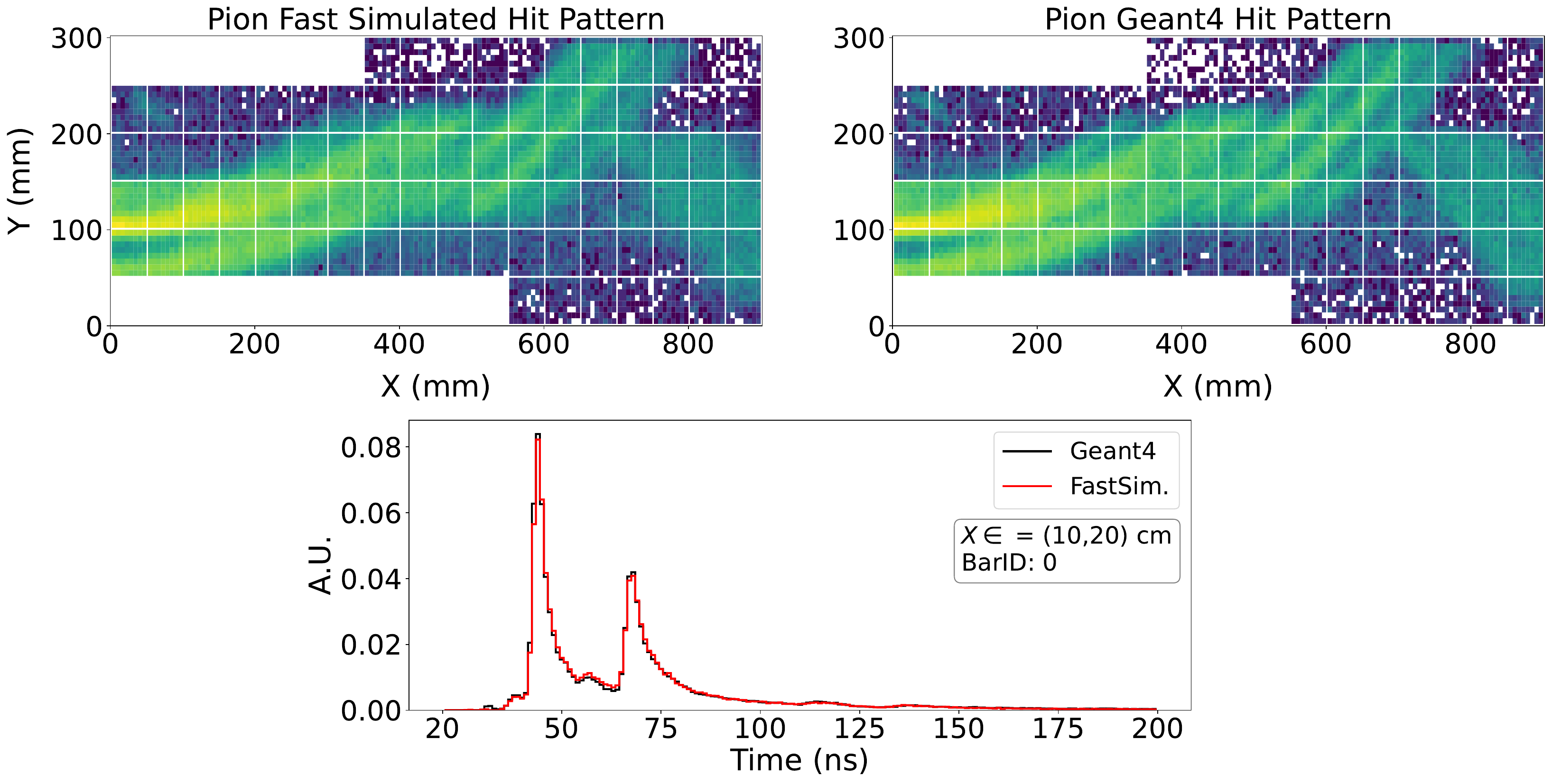}
    \includegraphics[width=0.49\textwidth]{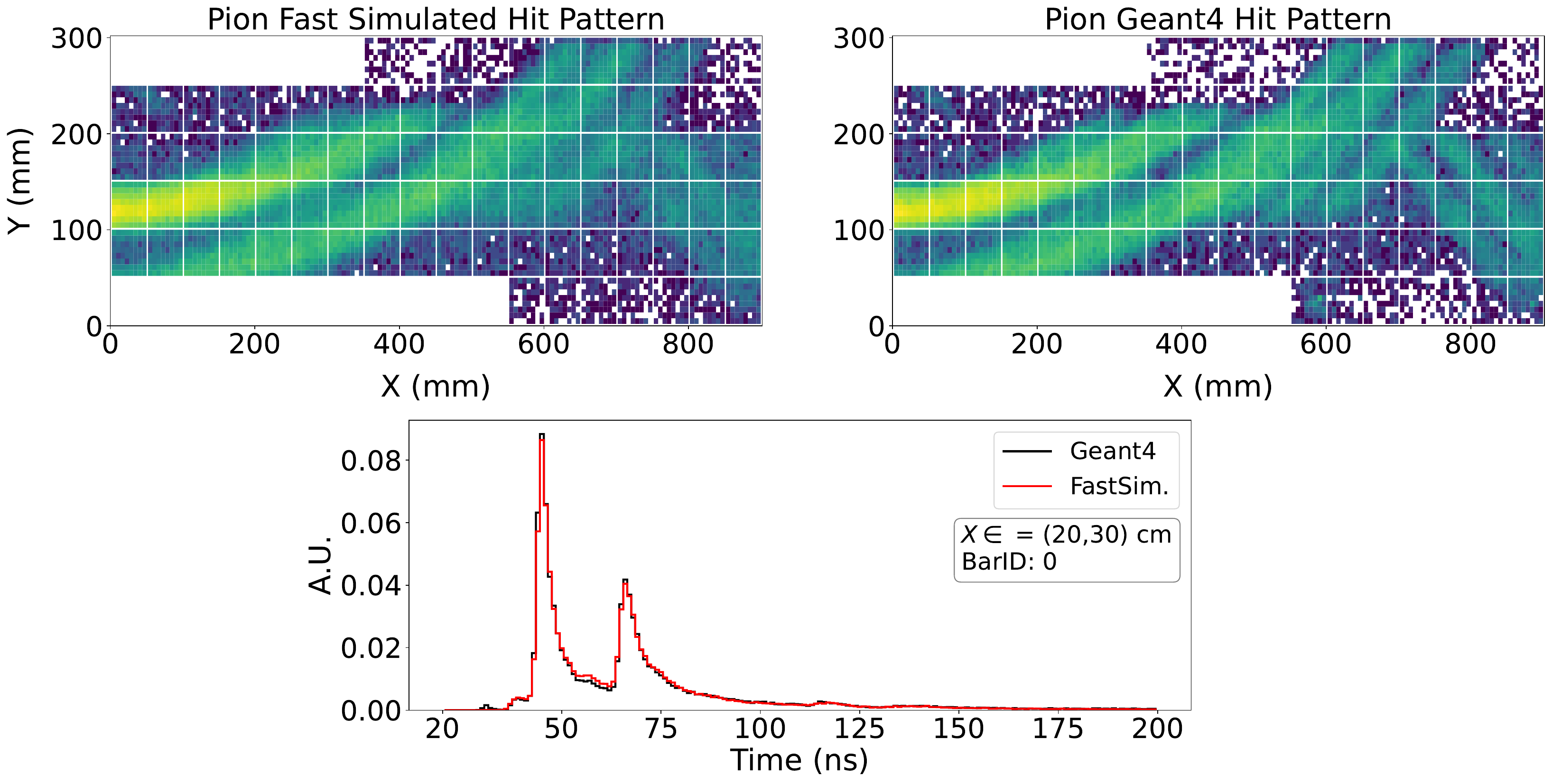}
    \caption{\textbf{Pion generations at bar 0}: Fast-simulated and \geant hit patterns shown across all positions along the bar face, scanned sequentially from left to right, top to bottom.}
    \label{fig:app_pion_bar0}
\end{figure}

\begin{figure}[h]
    \centering
    \includegraphics[width=0.49\textwidth]{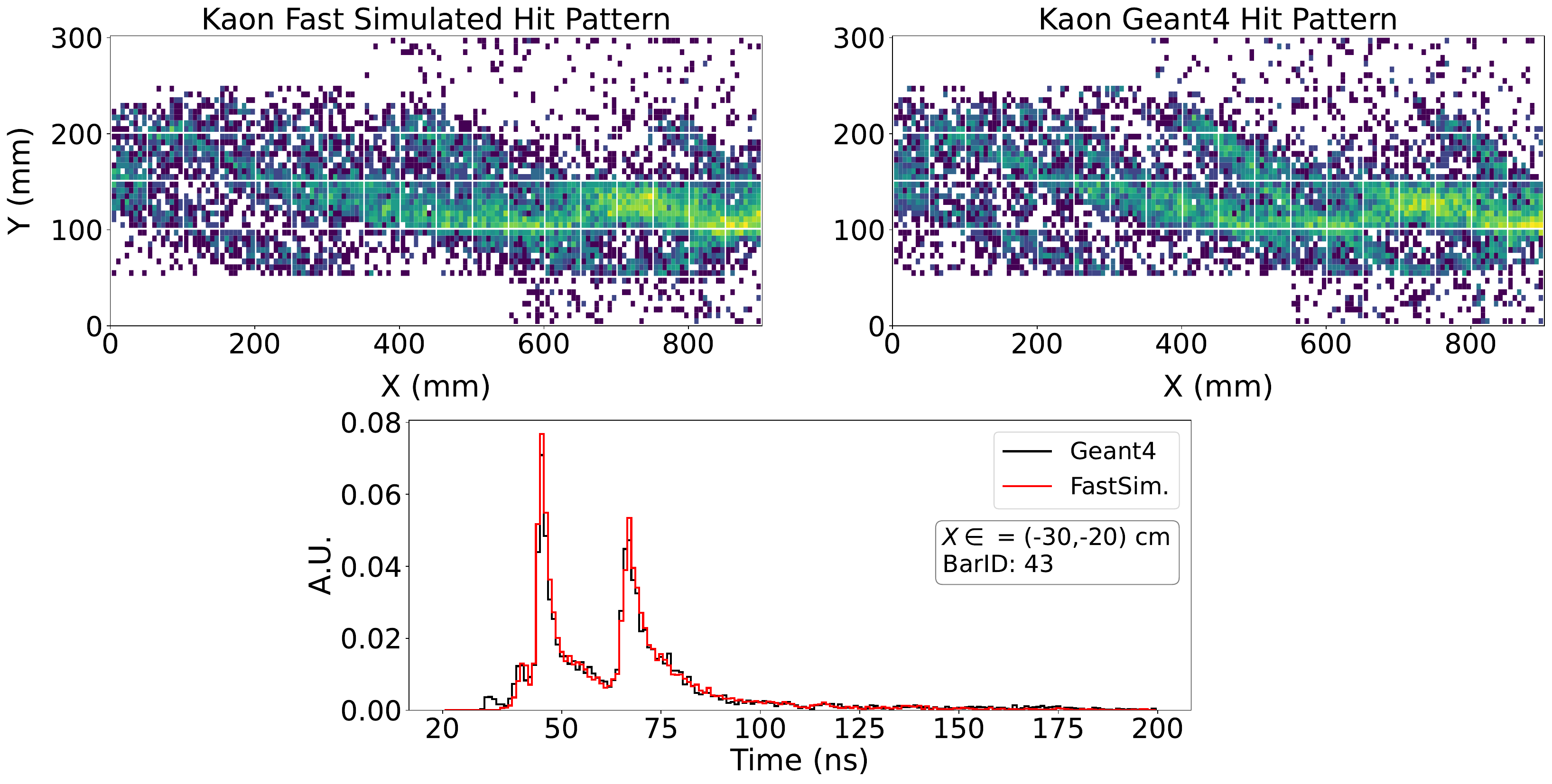}
    \includegraphics[width=0.49\textwidth]{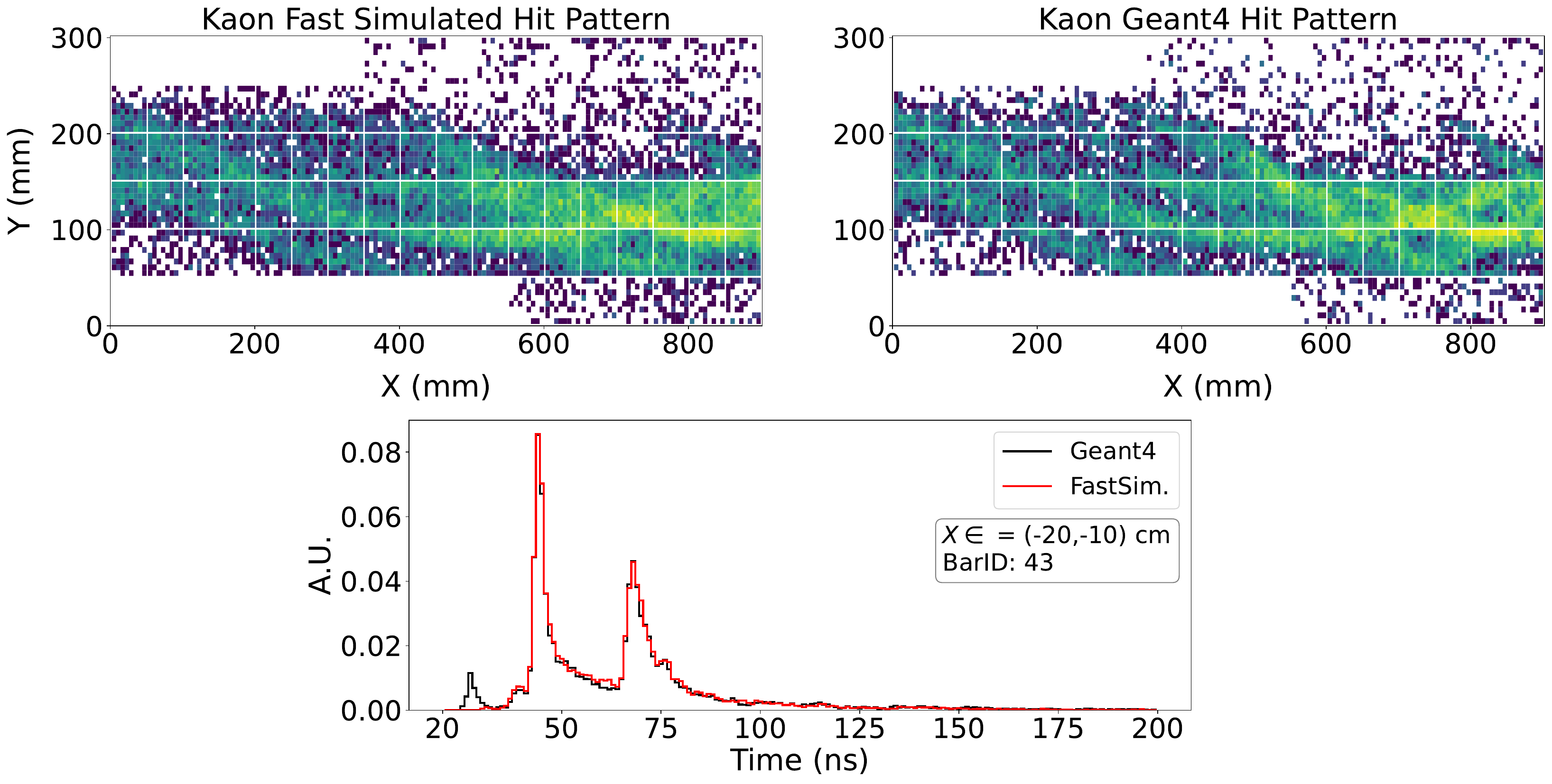}
    \includegraphics[width=0.49\textwidth]{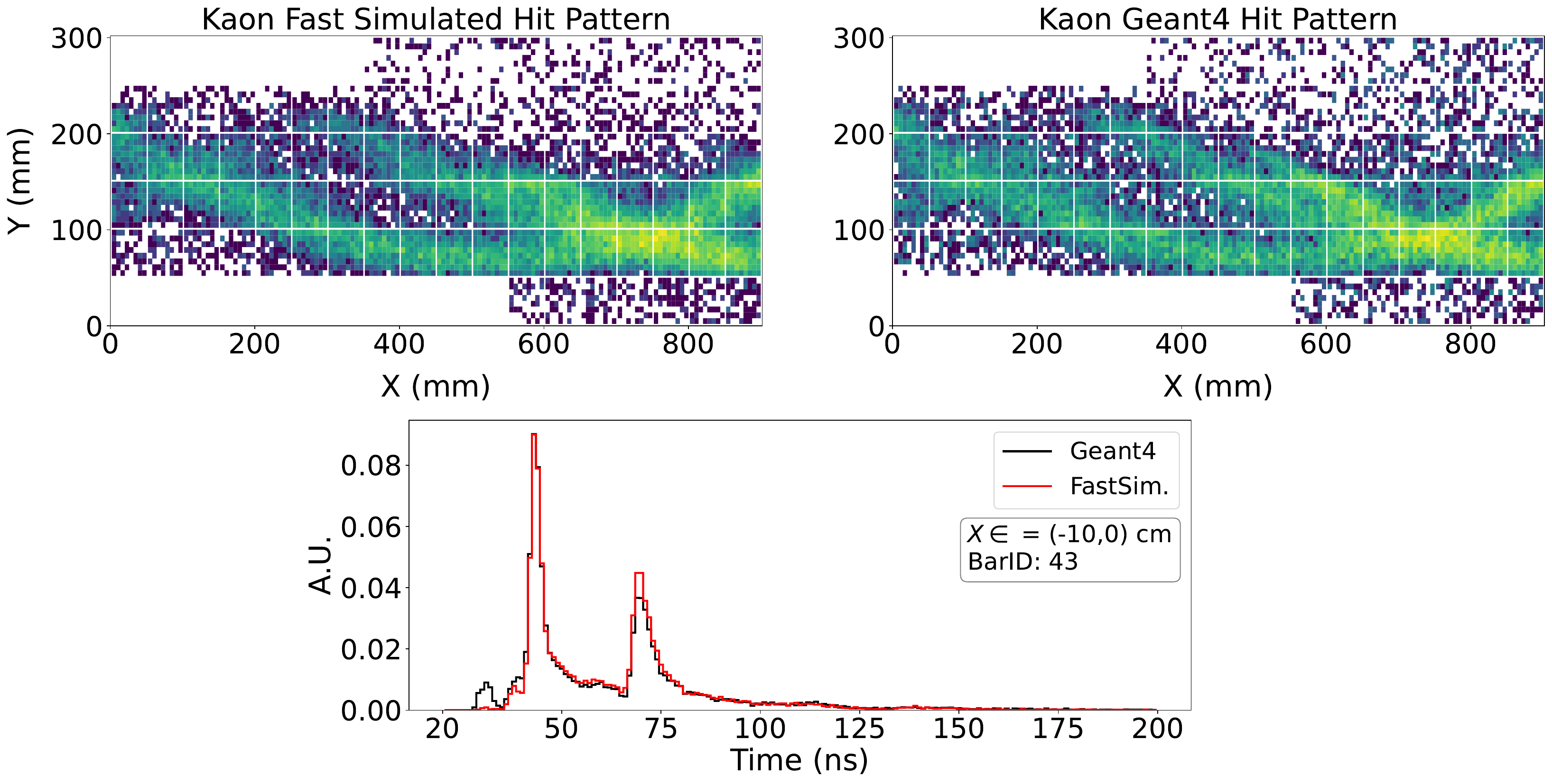}
    \includegraphics[width=0.49\textwidth]{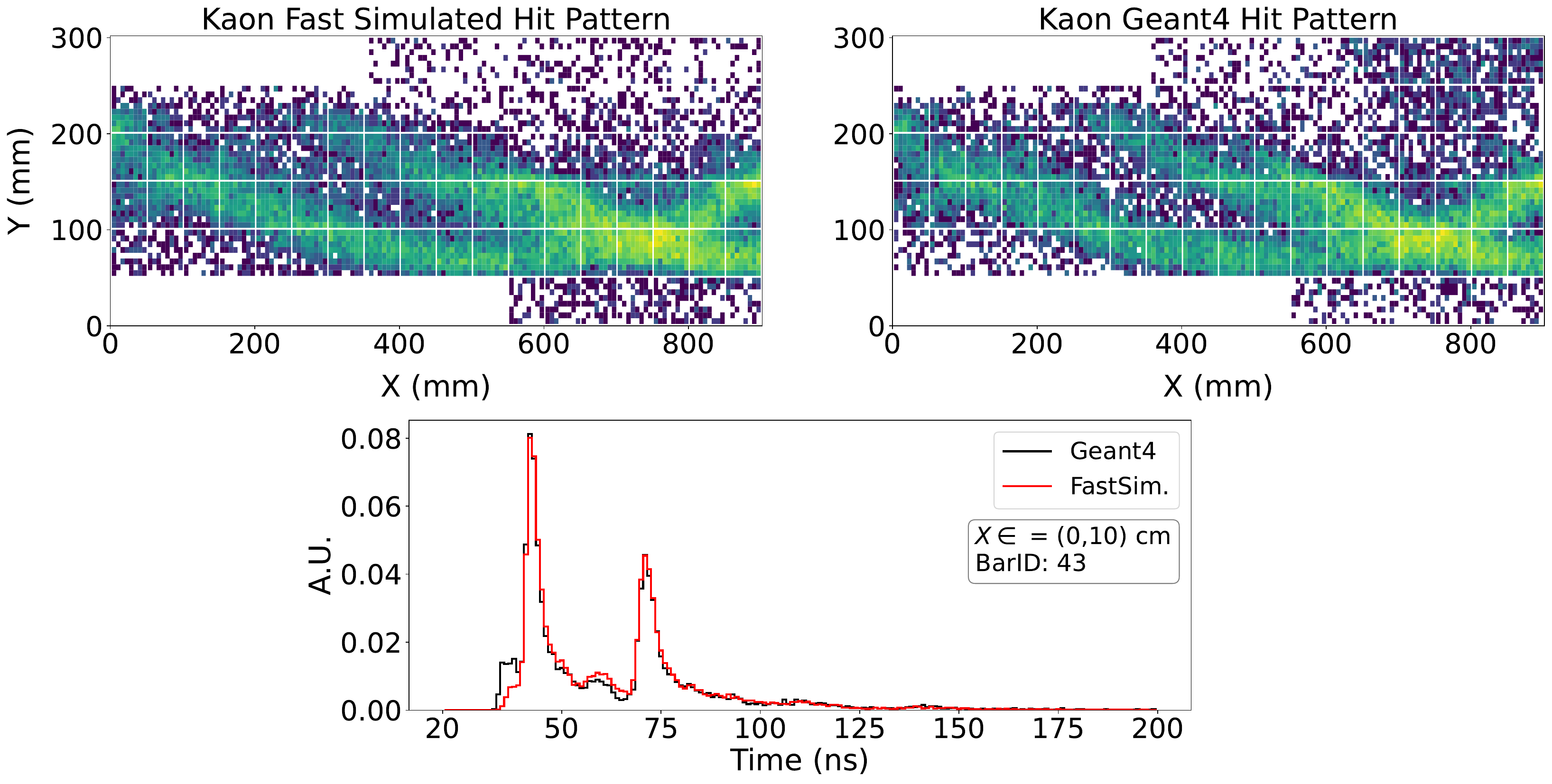}
    \includegraphics[width=0.49\textwidth]{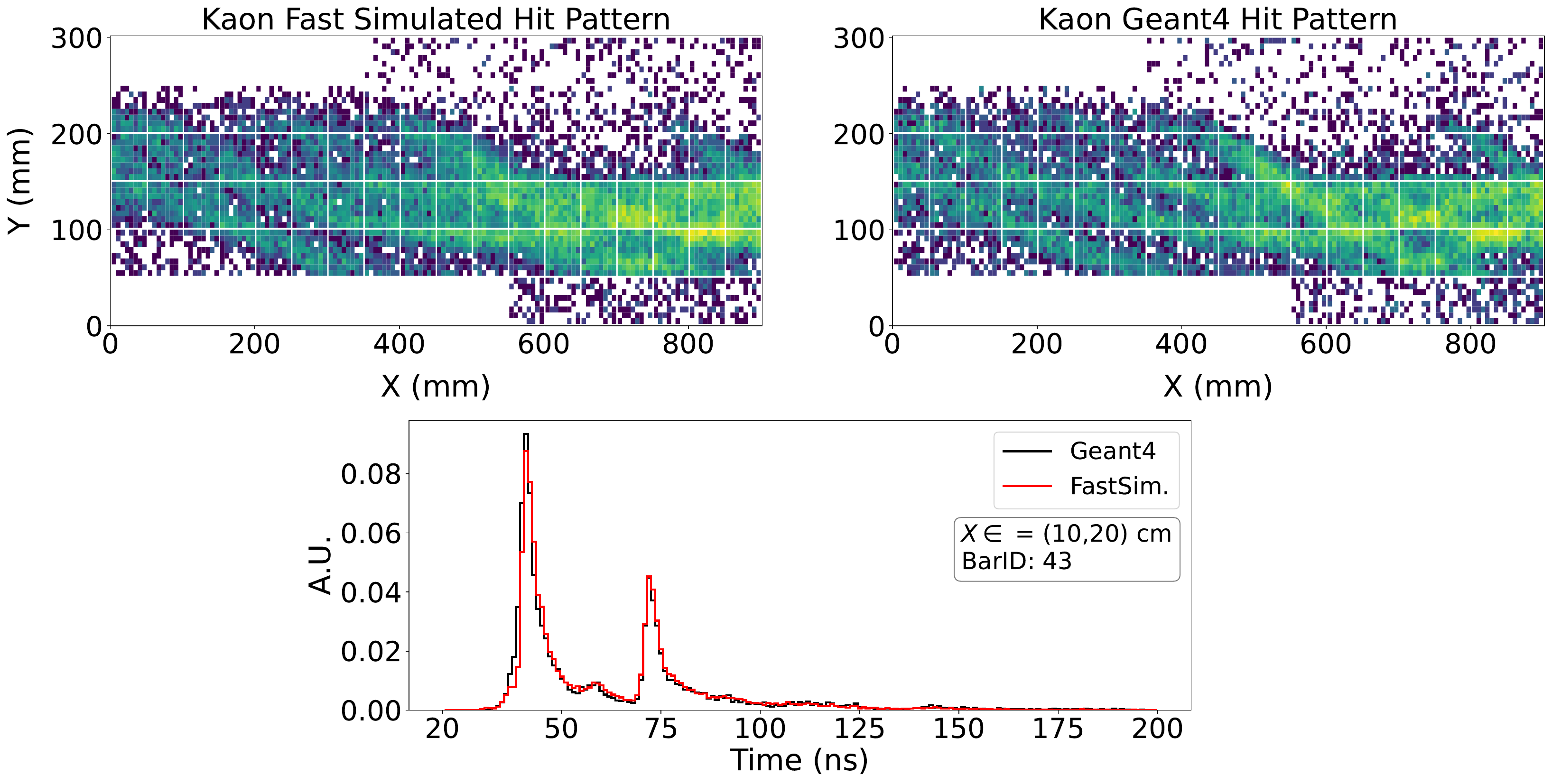}
    \includegraphics[width=0.49\textwidth]{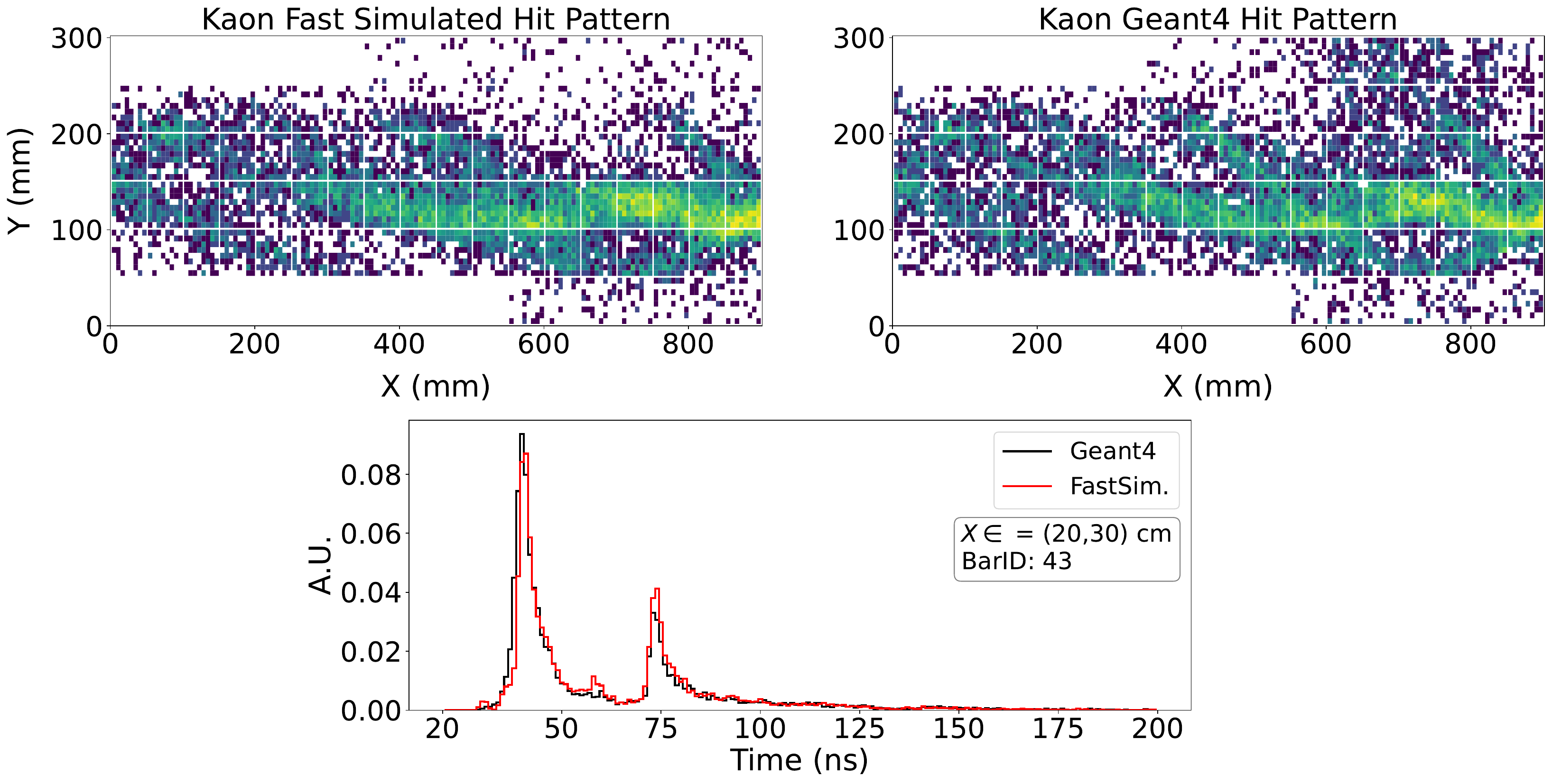}
    \caption{\textbf{Kaon generations at bar 43}: Fast-simulated and \geant hit patterns across all bar face positions. This sparsely populated training region highlights reduced fidelity relative to central regions.}
    \label{fig:app_kaon_bar43}
\end{figure}

\begin{figure}[h]
    \centering
    \includegraphics[width=0.49\textwidth]{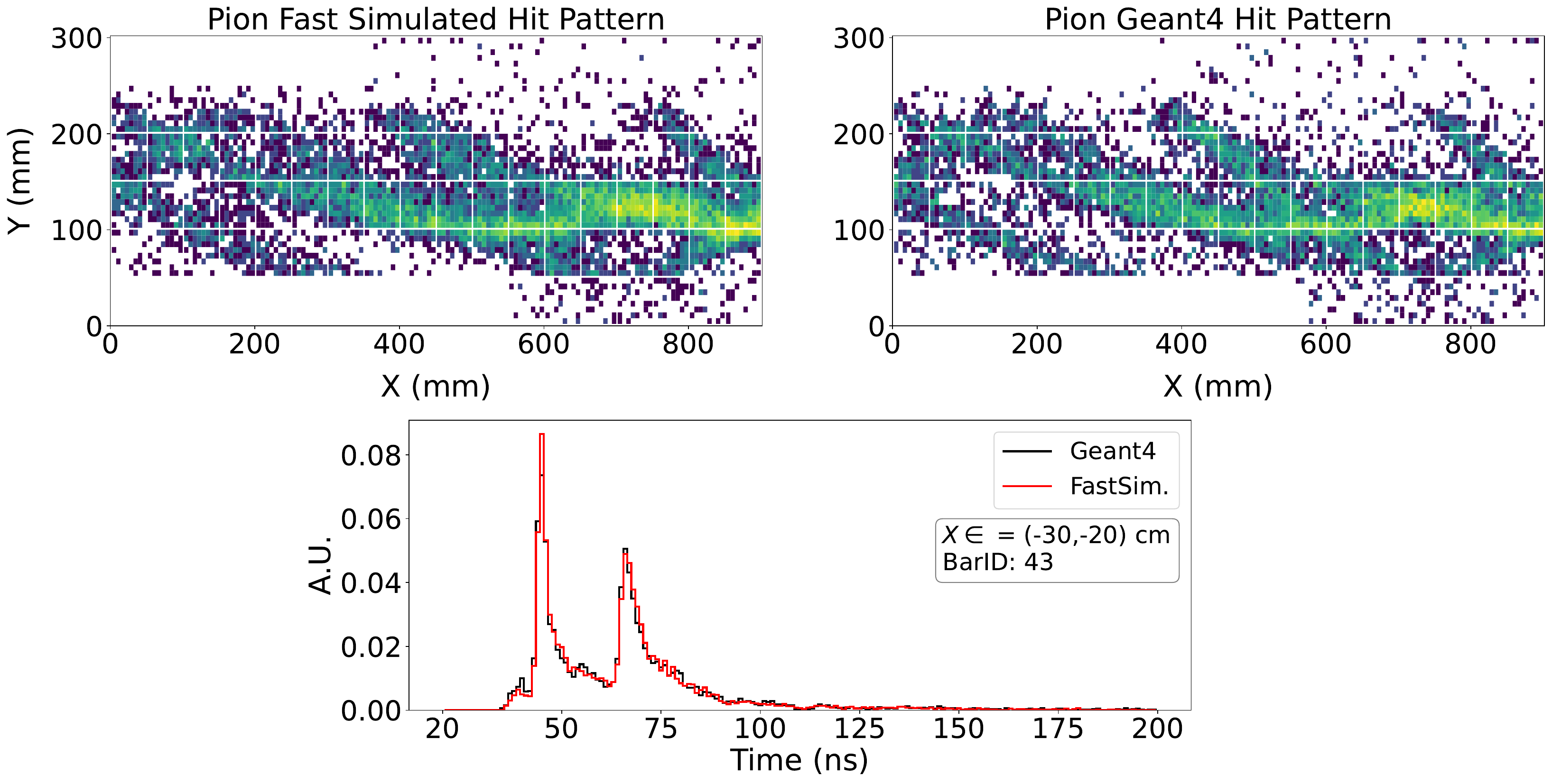}
    \includegraphics[width=0.49\textwidth]{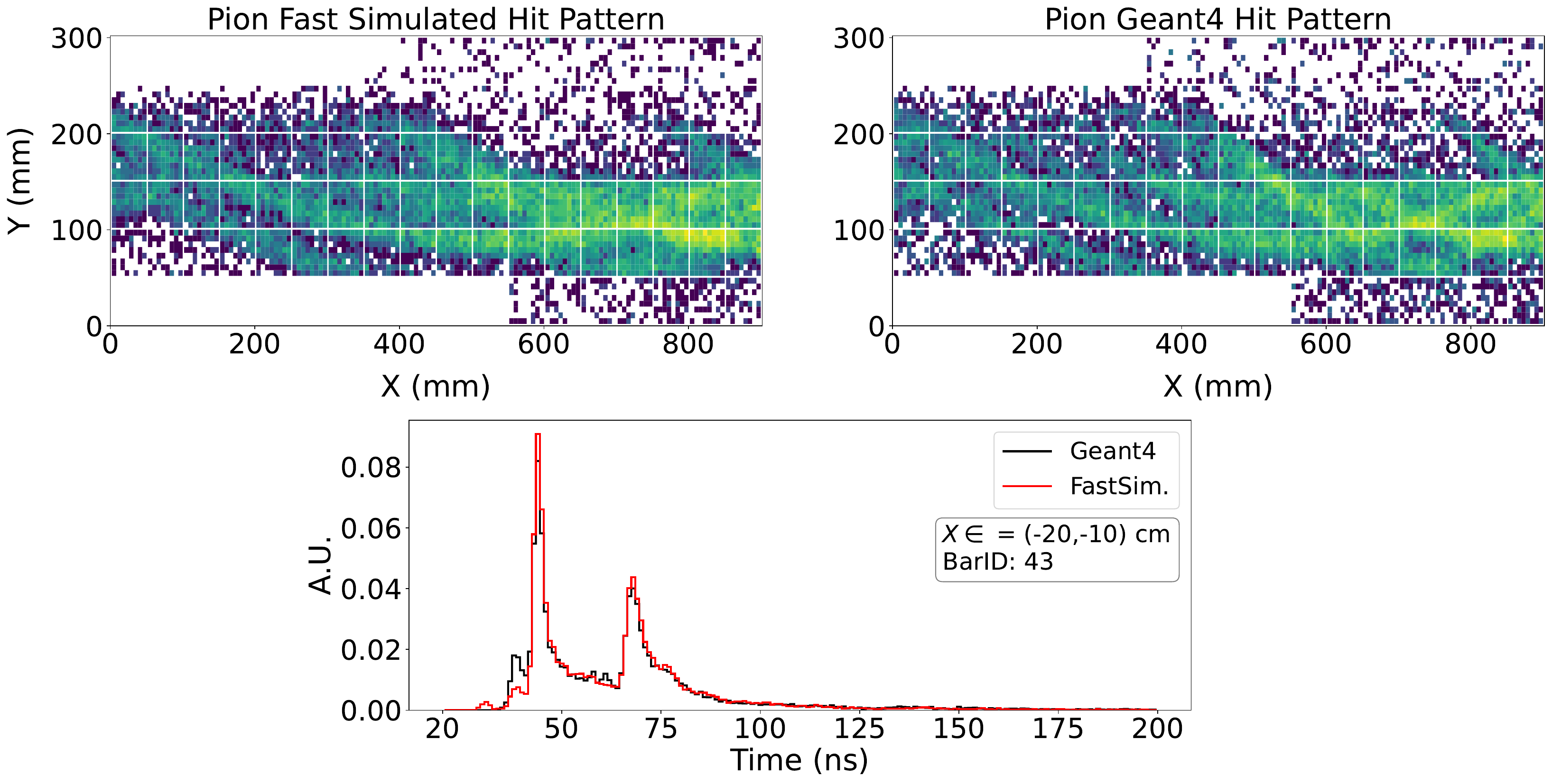}
    \includegraphics[width=0.49\textwidth]{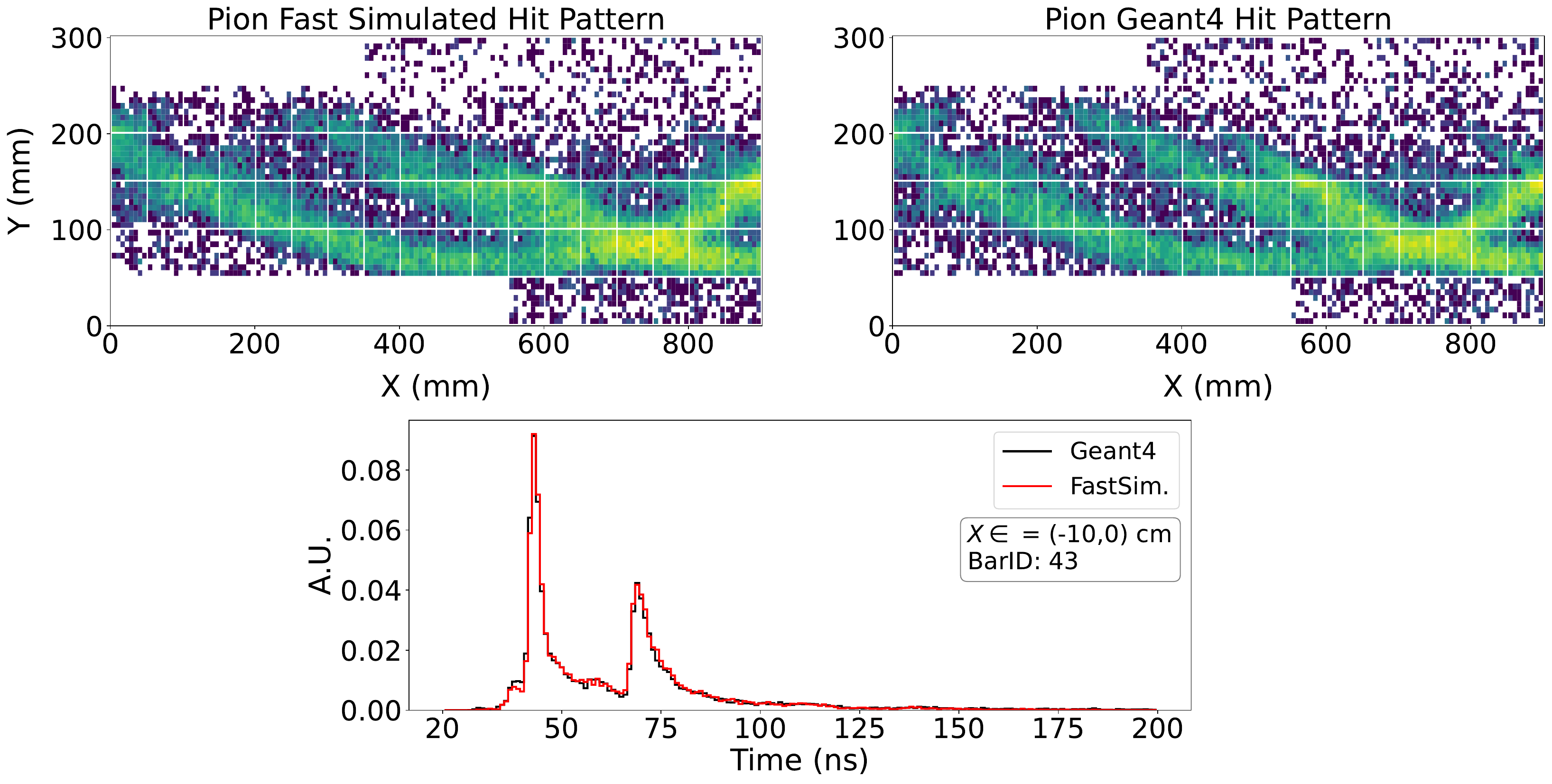}
    \includegraphics[width=0.49\textwidth]{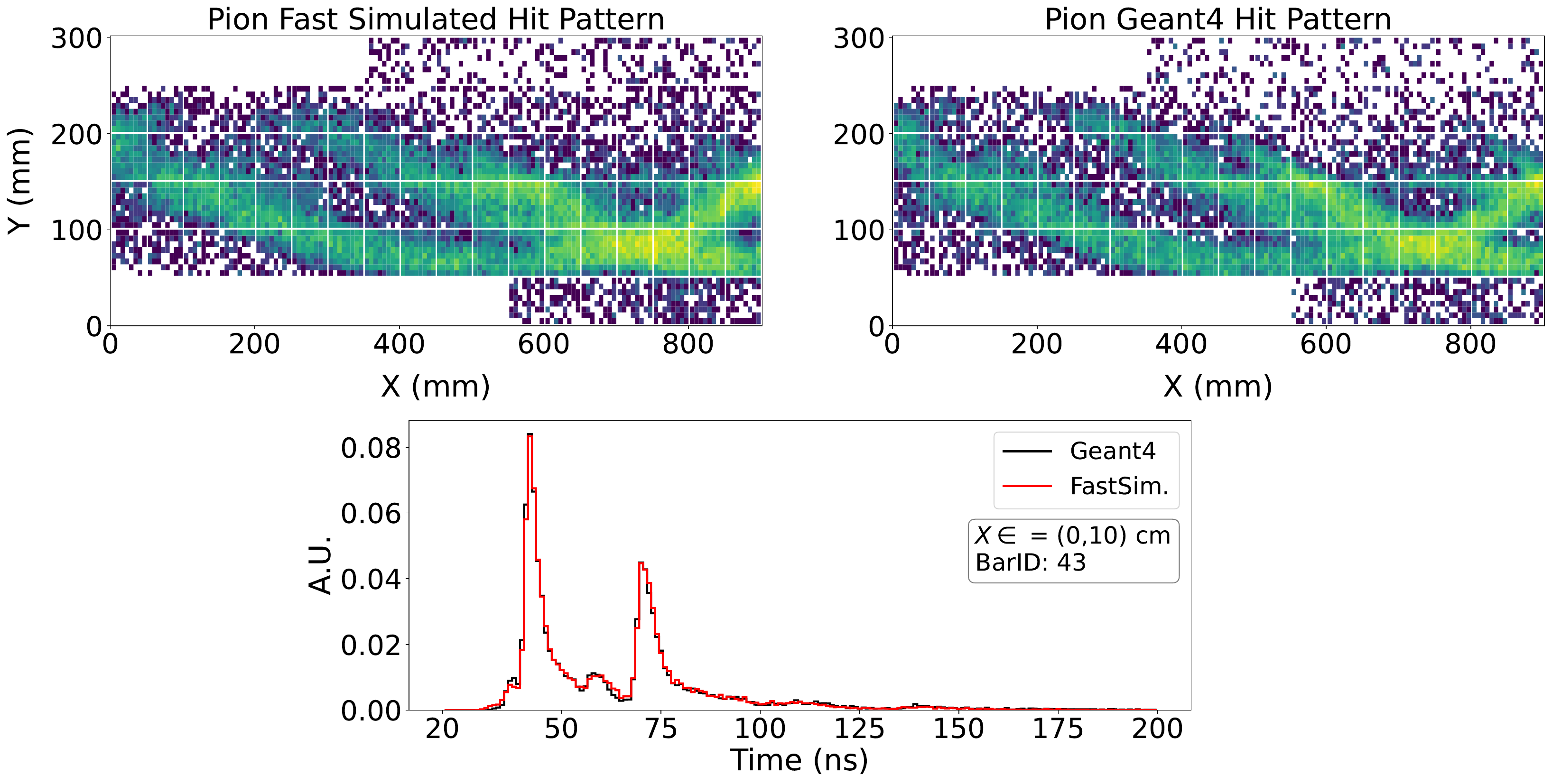}
    \includegraphics[width=0.49\textwidth]{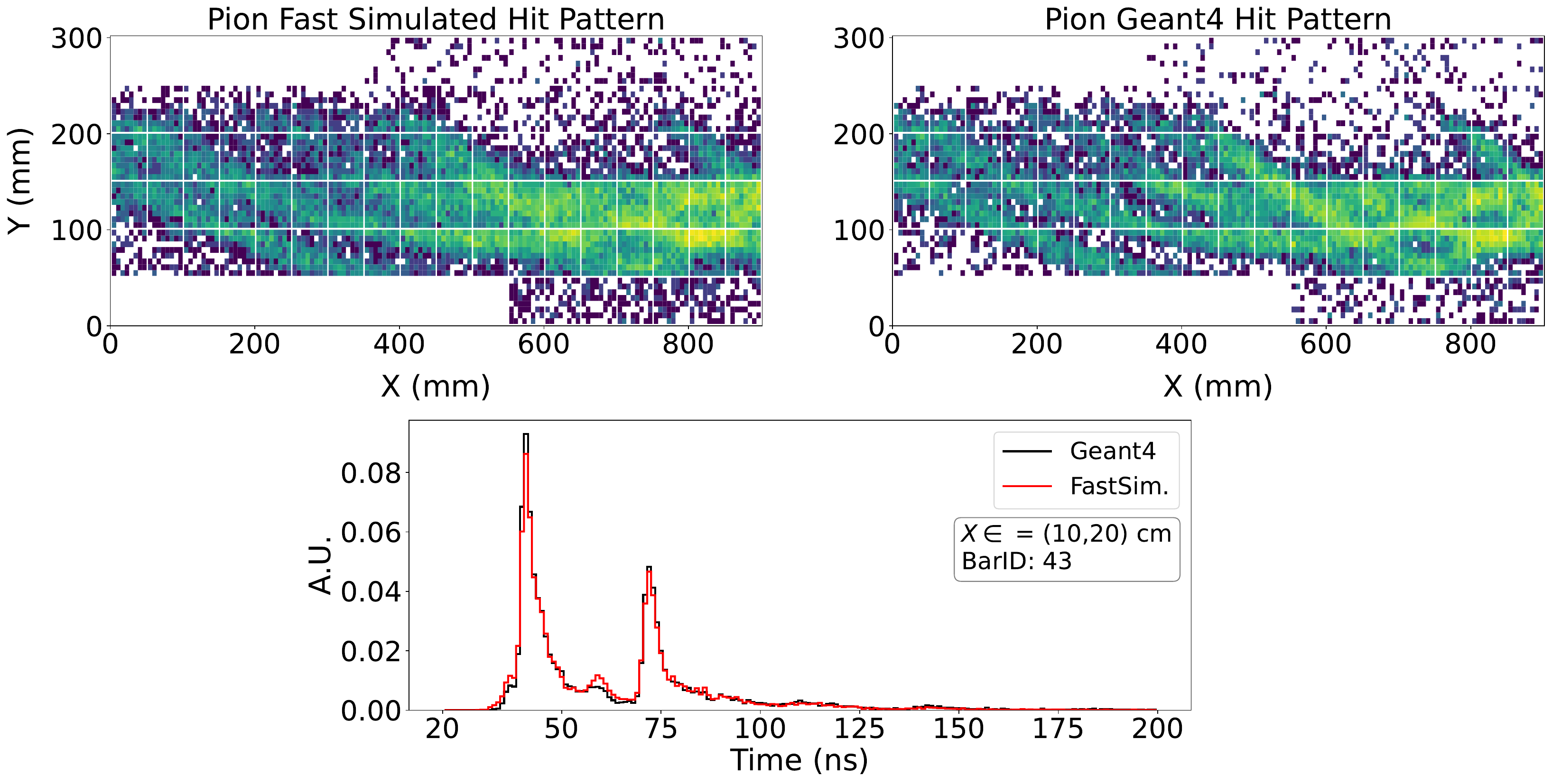}
    \includegraphics[width=0.49\textwidth]{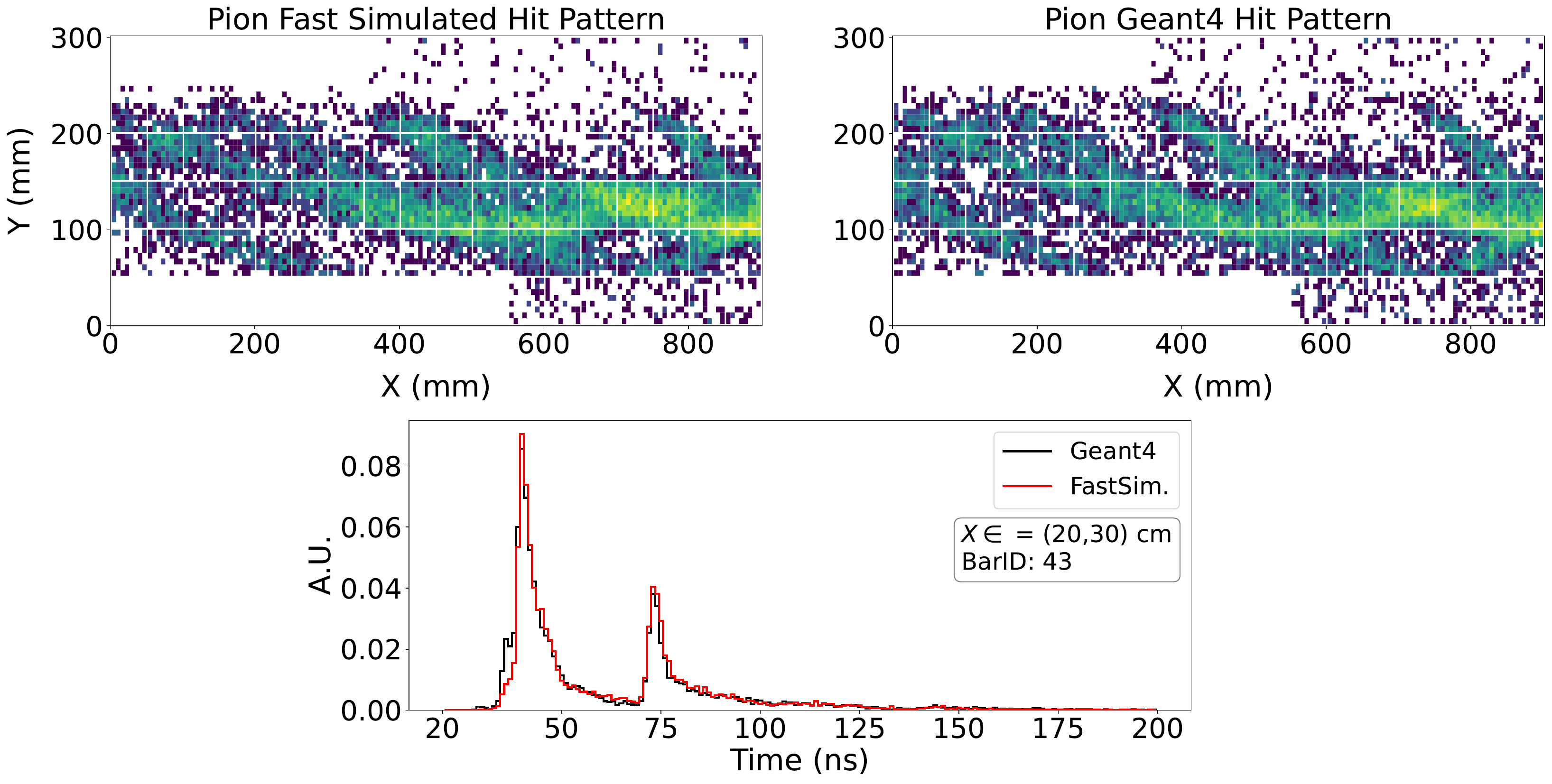}
    \caption{\textbf{Pion generations at bar 43}: Fast-simulated and \geant hit patterns across all bar face positions. Reduced training statistics in this region lead to visible degradation in generative fidelity.}
    \label{fig:app_pion_bar43}
\end{figure}

We provide additional generations in which all positions along the face of a given bar are scanned sequentially. Figures~\ref{fig:app_kaon_bar0} and~\ref{fig:app_pion_bar0} show kaon and pion generations alongside their ground truth counterparts at all face positions of bar 0, to be read left to right, row by row. Figures~\ref{fig:app_kaon_bar43} and~\ref{fig:app_pion_bar43} show the corresponding generations for bar 43, which represents a sparsely populated region of the phase space during training. While the model generally captures the underlying ring structure across all positions, performance in this region leaves room for improvement — a natural target for future pre-training with additional data.

